\documentclass{article}

\usepackage{graphicx}
\usepackage{color}

\makeatletter
\def\maxwidth{ %
  \ifdim\Gin@nat@width>\linewidth
    \linewidth
  \else
    \Gin@nat@width
  \fi
}
\makeatother

\definecolor{fgcolor}{rgb}{0.345, 0.345, 0.345}

\usepackage[utf8]{inputenc}
\usepackage{framed}
\makeatletter
 {\par\unskip\endMakeFramed%
 \at@end@of@kframe}
\makeatother

\definecolor{shadecolor}{rgb}{.97, .97, .97}
\definecolor{messagecolor}{rgb}{0, 0, 0}
\definecolor{warningcolor}{rgb}{1, 0, 1}
\definecolor{errorcolor}{rgb}{1, 0, 0}

\usepackage{alltt}
\usepackage[intoc]{nomencl}

\textfloatsep = 0.4in \addtocontents{toc}{\vspace{0.4in} \hfill
Page\endgraf} \addtocontents{lof}{\vspace{0.2in} \hspace{0.13in} \
Figure\hfill Page\endgraf} \addtocontents{lot}{\vspace{0.2in}
\hspace{0.13in} \ Table\hfill Page\endgraf}
\usepackage{float}
\usepackage{textcomp,bm}
\usepackage{array}
\usepackage{listings}
\usepackage{pdfpages}
\usepackage{setspace}
\usepackage{mathptmx}
\usepackage{capt-of}
\usepackage[table, svgnames]{xcolor}
\usepackage{colortbl}
\usepackage{graphicx}
\graphicspath{{fig/}}
\usepackage{amssymb, amsmath}
\usepackage{subfig}
\usepackage{wrapfig}
\usepackage{epsfig}
\usepackage{times}
\usepackage{authblk}
\usepackage{float}
\usepackage{epstopdf}
\usepackage{rotating}
\usepackage{bm}
\usepackage{comment}
\usepackage{lipsum}
\usepackage{rotating}
\usepackage{makeidx}
\usepackage{url}
\usepackage{multirow}
\usepackage{pdfpages,caption}
\usepackage{booktabs}
\usepackage[subfigure, titles]{tocloft}
\usepackage{caption}
\usepackage{acronym}
\usepackage{datetime}
\usepackage{pdfpages}
\usepackage{lmodern}
\usepackage[english]{babel}
\usepackage[autostyle]{csquotes}
\usepackage[linktocpage=true]{hyperref}
\usepackage{amsmath}
\usepackage[font=small,skip=0pt]{caption}
 \usepackage[margin=1in]{geometry}
\usepackage{algorithm}
\usepackage{lscape}
\usepackage{mdframed}
\usepackage[nottoc]{tocbibind}

\makenomenclature

\usepackage[nottoc]{tocbibind}
\IfFileExists{upquote.sty}{\usepackage{upquote}}{}

\usepackage{makecell}
\usepackage{titletoc}
\usepackage{appendix}
\usepackage{tikz}
\usetikzlibrary{decorations.pathmorphing} 
\usetikzlibrary{fit}
\usetikzlibrary{backgrounds}	
\usetikzlibrary{positioning}
\usetikzlibrary{shapes,decorations,arrows,calc,arrows.meta,fit,positioning}
\tikzset{
    -Latex,auto,node distance =1 cm and 1 cm,semithick,
    state/.style ={ellipse, draw, minimum width = 0.7 cm},
    point/.style = {circle, draw, inner sep=0.04cm,fill,node contents={}},
    bidirected/.style={Latex-Latex,dashed},
    el/.style = {inner sep=2pt, align=left, sloped}
     latentnode/.style={draw, minimum width=5mm, shape=circle, ultra thick, black},
  dagconn/.style={arrows=->, black, thick},
  plate/.style={draw, shape=rectangle, rounded corners=0.5ex, thick,
    minimum width=3.1cm, text width=3.1cm, align=right, inner sep=10pt, inner ysep=10pt, 
    append after command={node[above left= 3pt of \tikzlastnode.south east] {#1}}}
}

\makenomenclature

\title{A Bayesian hierarchical small-area population model accounting for data source specific methodologies from American Community Survey, Population Estimates Program, and Decennial Census data\\
}

\author[1]{Emily N. Peterson}
\author[2]{Rachel C. Nethery}
\author[3]{Tullia Padellini}
\author[2]{Jarvis T. Chen}
\author[2]{Brent A. Coull}
\author[3]{Fr\'ed\'eric B. Piel}
\author[4]{Jon Wakefield}
\author[3]{Marta Blangiardo}
\author[1]{Lance A. Waller}

\affil[1]{Department of Biostatistics and Bioinformatics, Rollins School of Public Health, Emory University, Atlanta GA USA}
\affil[2]{Department of Biostatistics, Harvard TH Chan School of Public Health, Boston MA USA}
\affil[3]{Department of Epidemiology and Biostatistics, School of Public Health, Imperial College London, London UK}
\affil[4]{Department of Biostatistics, School of Public Health, University of Washington, Seattle WA USA}

\usepackage[nottoc]{tocbibind}
\IfFileExists{upquote.sty}{\usepackage{upquote}}{}

\begin{document}
\pagenumbering{alph}


\pagenumbering{roman} \setcounter{page}{1}
\captionsetup{font=small,skip=0pt}
\maketitle

\tableofcontents

\begingroup
\setlength{\parskip}{1\baselineskip}
\printnomenclature
\newpage
\endgroup

\normalsize

\pagenumbering{arabic}
\setcounter{page}{1}
\doublespacing

\begin{abstract}
\singlespace{
  Small area estimates of population are necessary for many epidemiological studies, yet their quality and accuracy are often not assessed. In the United States, small area estimates of population counts are published by the United States Census Bureau (USCB) in the form of the Decennial census counts, Intercensal population projections (PEP), and American Community Survey (ACS) estimates. Although there are significant relationships between these data sources, there are important contrasts in data collection and processing methodologies, such that each set of estimates may be subject to different sources and magnitudes of error. Additionally, these data sources do not report identical small area population counts due to post-survey adjustments specific to each data source. Resulting small area disease/mortality rates may differ depending on which data source is used for population counts (denominator data). To accurately capture annual small area population counts, and associated uncertainties, we present a Bayesian population model (B-Pop), which fuses information from all three USCB sources, accounting for data source specific methodologies and associated errors. We produce comprehensive small area race-stratified estimates of the true population, and associated uncertainties, given the observed trends in all three USCB population estimates. The main features of our framework are: 1) a single model integrating multiple data sources, 2) accounting for data source specific data generating mechanisms, and specifically accounting for data source specific errors, and 3) prediction of estimates for years without USCB reported data. We focus our study on the 159 counties of Georgia, and produce estimates for years 2005-2021. We compare B-Pop population estimates to decennial census counts, ACS 5-year estimates, and PEP annual counts. Additionally, we illustrate and explain the differences in estimates of data-source specific errors. We show, through simulation,  the implications of the differing degrees of data-source specific errors on the associated uncertainties of small area population estimates. Lastly, we compare model predictive performance using validation exercises. Our Bayesian framework accounts for differing data generating mechanisms and data source specific errors in the estimation of small area population estimates, with the ability to predict to population-periods without data.
}
    \bigskip
    
    \noindent
    \textbf{Acknowledgements:} The authors are very grateful to all the members of the Spatial Uncertainty Research Team, from the Emory Rollins School of Public Health, Harvard T.H. Chan School of Public Health, and the UK Small Area Health Statistics Unit (SAHSU) at Imperial College London's School of Public Health, for the support of this work.
    \bigskip
    
    \noindent
    \textbf{Funding:} This work, led by Lance A. Waller, and was funded by the US National Institutes for Health (NIH) (\#RO1HD092580). FBP acknowledges support from Health Data Research UK(HDR UK) and the UK National Institute for Health Research (NIHR) Imperial Biomedical Research Centre. FBP is a member of the Health Protection Research Units in Chemical and Radiation Threats and Hazards, and in Environmental Exposures and Health, which are partnerships between Public Health England and Imperial College London funded by the UK NIHR.
    \bigskip
    
\end{abstract}

\section{Introduction}
Small area estimates are routinely used to track local population trends in disease and mortality rates. Specifically, health research uses population data as denominators in estimation of disease and/or mortality rates (Nethery, R. et al (2021)).  In the United States, small area estimates of population counts are published by the United States Census Bureau (USCB) in the form of the Decennial Census, Intercensal population projections (PEP), and the American Community Survey (ACS). In studies tracking small area disease rates over time, researchers often use ACS small-area multi-year estimates to capture annual trends in the at risk population. In addition, these population estimates are used to obtain up-to-date (i.e., current year) annual small area disease or mortality rates. Although each source reports population data at the county level, there are important distinctions in data methodologies, and data availability, based on geographic unit, i.e., counties and states have higher data availability across the different sources, compared to census tracts and blocks. In addition, the different methodologies yield population estimates, which capture incomparable data quantities. As a result, small area disease or mortality rates can differ depending on which source of population data are used as denominator data (Nethery, R. et al (2021)). \\

Although rarely acknowledged, each of these different population data sources has limitations, of which data users must be aware. First, there are delays in reporting, i.e., the most recent available data for 2021 is the PEP 2020 population estimates, and the most comprehensive decennial data are only available every 10 years. Therefore, USCB data do not capture the most recent year population estimates, and the population-at-risk denominators, used in small area health statistics, may be from earlier years. Secondly, small area annual intercensal counts are not reported for census tract or block groups within counties. This gap is particularly important when exploring small-area patterns in population sub-groups (e.g., race, socio-economic status). Lastly, small area (census tracts and block groups) population estimates are reported by ACS in the form of 5-year period estimates. Therefore, users may introduce bias in small area disease surveillance with the use of delayed and/or non-annual small area population data. The biases that can be induced in disease mapping models due to inaccurate population size estimates were recently explored in Nethery, R. et al (2021). \\

Most spatial analyses of disease rates treat population estimates as fixed, known quantities, and do not account for measurement error (uncertainty) associated with the reported estimates, in part due to the absence of reported uncertainties with many USCB products. However, not accounting for uncertainty in population data may lead to bias in small area estimates, as well as inaccurate estimation of the uncertainty associated with these estimates. The motivation for this work is to build upon the rich data and methods, reported per USCB data sources, to produce annual race-stratified small area (county level) population size estimates that are accurate, comprehensive, and fully compatible across time and space, and that are accompanied by representative uncertainty estimates. These improved population size estimates, and uncertainties, will allow for more accurate characterization of small area disease and mortality trends.\\

There are important contrasts in data collection, and processing, methodologies between the different USCB programs, such that reported population estimates from the different programs may be subject to different sources, and magnitudes, of error and different post-survey adjustments. These differences in estimates, and associated errors, present significant challenges when combining data across sources, and quantifying uncertainties. Our aim is to present a Bayesian Hierarchical Population Estimation (B-Pop) model, which fuses information from all three USCB population data sources in a principled way, accounting for documented source-specific errors, and processing procedures, and  produces unified, comprehensive small area race-stratified population estimates accompanied by uncertainties. In addition, the B-Pop model allows for prediction of small area population estimates to years without data, i.e., more recent years in which ACS, PEP, and decennial data are unavailable. We also show how annual county specific estimates, obtained from the B-Pop model, can be used to impute smaller area 1-year population counts (i.e., for census tracts), which are not currently available per USCB programs. The resulting population estimates, and uncertainties, across years, provide small area annual race-stratified population counts (and associated errors) for counties and census tracts, predicted up to the most recent years, and as such can be used as denominator estimates within a larger disease mapping framework.\\ 

Our objective is to account for data source specific errors associated with different data methodologies. First we address non-sampling error, i.e., error not attributable to a sample survey design, which impacts all USCB data sources. The lack of data to directly inform non-sampling error estimates requires that we use a model-based approach. In addition to non-sampling error, the careful design of the ACS allows specific calculation of associated sampling errors. Previous work compared total ACS reported uncertainties with those of the 2000 decennial census long form data, and showed that ACS margins of error were, on average, $75\%$ larger than those corresponding to the 2000 decennial census (Spielman, S. et al. (2014)). These results indicate that, for the smallest geographies, ACS population estimates suffer from substantial uncertainty, compared to population sizes reported by the decennial census, due to sampling uncertainty (design-based sampling error). Additionally, the ACS survey methods yield a smaller sample compared to the complete enumeration of the decennial census (Spielman, S. et al. (2014); Starsinic, M. and Tersine Jr., A. (2007)). The authors note that the key reasons for the ACS's increased uncertainty, compared with the 2000 decennial census long form, were its sampling of higher geographic resolution, i.e., sampling of smaller local sample sizes will increase errors surrounding population estimates, and an increase in the variability in survey weights. Additionally, Spielman \textit{et al}. showed the existence of spatial patterns in associated errors, and that ACS reported error varied by demographic sub-populations (i.e., heterogeneity of the population).  These differences in levels of data-source specific uncertainty, highlight the need to account for data-source specific errors associated with population estimates, specifically in the context of quantifying small area sub-population health disparities
(Spielman, S., et al. (2014); Spielman, S., and Folch, D. (2015)). \\

To incorporate data source specific error terms within B-Pop, we extend the classical additive error model framework, in which observed data are assumed unbiased, and equal to the sum of the true counts plus a stochastic error term (Carroll, R.J. et al. (2006)). We build on an extension proposed by Riebler \textit{et al}. (2014) to relax the assumption that ACS data are unbiased for the true population counts. Specifically, in our approach, the stochastic error term is broken down into two unique additive components: (1) a county-race specific model-based non-sampling error term, and (2) an observed county-year-race specific noise component introduced by ACS sampling schemes, a design-based sampling error.\\

Small area estimation (SAE) is a hybrid between design-based inference, and model-based estimates. To produce local population estimates that account for survey weighting schemes, several studies propose adjusted design-weight based estimators in combination with a Bayesian smoothing model in the assessment of complex health-related survey data. The design weights are incorporated into the likelihood, and reduce variance through spatial smoothing to produce small area estimates, which account for sampling schemes, and reduce variability in highly variable small sample areas (Mercer, L. et al. (2014); Checn, C. et al. (2014); Mercer, L. et al. (2015)). A limitation of the proposed methods is how to account for uncertainties in the population-at-risk, specifically, how to account for data source specific errors when using population count data gathered from different sources. Based on findings of Spielman \textit{et al}. (2014) it is necessary to acknowledge the differential sampling error introduced by ACS data, and to assess how we account for these errors in the estimation of population counts using a model-based approach. As such, our approach combines a design-weight based estimator, and a Bayesian smoothing approach, as well as incorporating measurement error parameters, and their respective models, within a unified statistical model framework. Bradley \textit{et al.} (2015) presented a hierarchical model to perform spatio-temporal change of support for ACS multi-year survey data, accounting for sampling errors. They used a mixed effects model, which provides multi-resolution estimates through basis function aggregation in space and time with the aim of obtaining population estimates of different space and time resolutions, based on reported multi-year ACS period estimates, which required a more detailed change of support procedure (Bradley, J. et al (2015)). In contrast, our aim is to account for the overlapping nature of ACS multi-year survey estimates, and to use ACS, PEP, and decennial census to inform our estimates.\\  

In summary, when fusing information from the three USCB data sources, the B-Pop model: (1) incorporates multiple data models to account for the data source specific data generating mechanisms, and associated errors, i.e., the processes by which population estimates are derived across the different sources, (2) estimates county, and census tract specific, populations by White only, and Black only sub-groups to demonstrate the robust use of B-Pop to capture demographically defined sub-populations, and (3) implements a  Bayesian hierarchical random walk (RW2) process model to predict estimates of true small area population counts to years without data  (Shumway, R.H. and Stoffer, D.S. (2006)) We apply the B-Pop model to estimate county and census tract level race-stratified predicted population estimates and uncertainties for the state of Georgia (GA), over the period 2005-2021. We note B-Pop has broad applicability across geographies and demographic groups that have multiple sources of small area population data available.\\

\begin{figure}[H]
\centering
    \resizebox{\textwidth}{!}{%
\begin{tikzpicture}
  \node[state, rectangle,fill=red!10] (cendat) at (-4,6) {2.1 Decennial census};
   
   \node[state, rectangle,fill=red!10] (acsdat) at (-4,5) {2.2 American Community Survey data};

  \node[state, rectangle,fill=red!10] (pepdat) at (-4,4) {2.3 Intercensal Projections};
  
    \node[state, rectangle,fill=red!10] (dataava) at (-4,3) {2.4 Data Availability};
  \node[plate={2 Data}, blue,inner sep=20pt, fit=(acsdat) (cendat) (pepdat)(dataava) ] (plate1) {};

   \node[state, text width = 5cm, rectangle,fill=red!10] (mmeth) at (5,8) {3.1 Summary of methods};
   
   \node[state, text width = 5cm, rectangle,fill=red!10] (vars) at (5,6) {3.2 Description of sources of data variability:\\
   Description of data source specific errors};
   
  \node[state, text width = 5cm, rectangle,fill=red!10] (dm) at (5,4) {3.3 Data Models:\\
Description of likelihood functions and prior assumptions};
  
  \node[state, text width = 5cm, rectangle,fill=red!10] (mod) at (5,1.8) {3.4 Modeling unobserved population counts:\\
  Estimation of the true population with a random-walk model};
  
       \node[state, rectangle,fill=red!10] (tract) at (5,0) {3.5 Derivation of tract estimates};
       
          \node[state, text width = 5cm, rectangle,fill=red!10] (val) at (5,-1.3){3.6 Model Validation:\\
          Assessment of model predictive performance};

 \node[plate={3 Methods},  blue,inner sep=20pt, fit=(mmeth) (dm) (vars) (mod) (tract) (val) ] (plate2) {};

   \node[state, rectangle,fill=red!10] (vars) at (14,7) {4.1 Global estimates};

  \node[state, text width = 5cm, rectangle,fill=red!10] (nse) at (14,5.4) {4.2 County estimates of non-sampling errors:\\
  Comparison to population size and reported sampling errors};
  
    \node[state, rectangle,text width = 5cm,fill=red!10] (county) at (14,3.5) {4.3 County population size estimates and uncertainties};
    
       \node[state, rectangle,fill=red!10] (valres) at (14,2.3) {4.4 Validation results};
       
          \node[state, text width = 5cm, rectangle,fill=red!10] (map) at (14,1){4.5 Comparison of population estimates between USCB and B-Pop};
         
 \node[plate={4 Results},  blue,inner sep=20pt, fit= (vars) (nse) (county) (valres) (map) ] (plate3) {};
\end{tikzpicture}
}
\hspace{0.2cm}

\caption{Outline of the structure of sections and descriptions.  }
\label{fig:flow}
\end{figure}
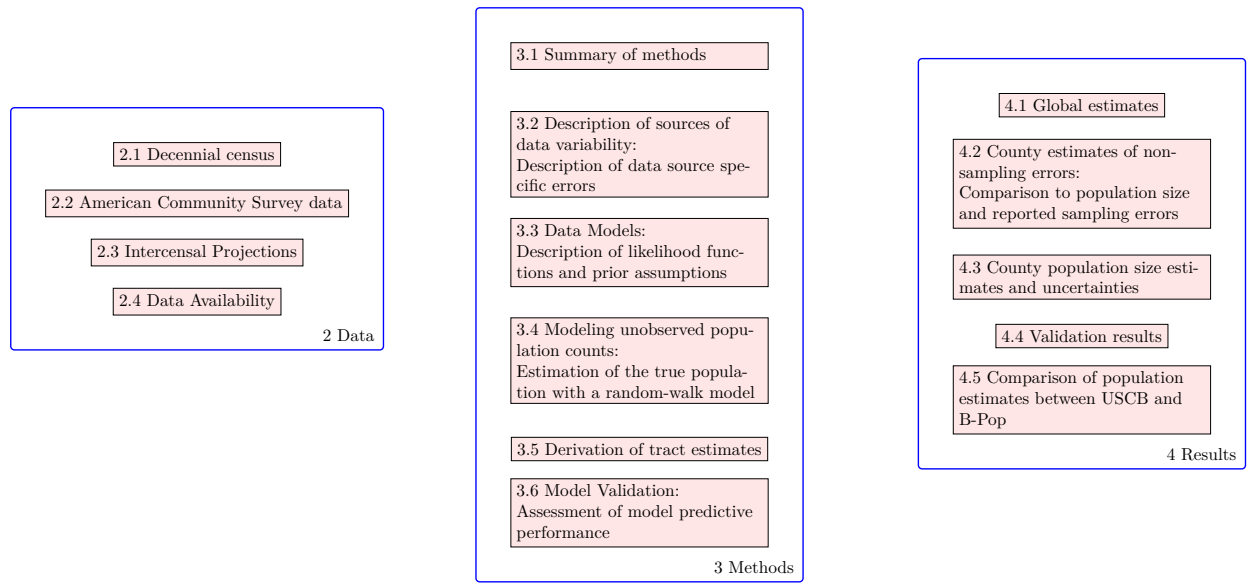

Figure \ref{fig:flow} presents a guide to the structure of this paper. The following section (Section \ref{sec-data}) describes the different USCB data products, and their availability. Section \ref{sec:mod} describes our approach to construct county/census tract, and race-stratified population estimates, and gives a detailed description of B-Pop components. Specifically, we describe the data source specific likelihood functions, and modeling of associated errors. We describe the modeling of unobserved population counts, which extends to prediction of estimates for population-periods without data, including smaller population of census tracts. Lastly, we compare model predictive performance using validation studies. Section \ref{sec:res} shows our results for county-year-race/tract-year-race estimates and associated model-based uncertainty estimates for Georgia between 2005-2021, detailing comparison of sampling and non-sampling estimates, global and county specific temporal trends, and results from our validation study. Lastly, in Section 5, we discuss the implications of our results for further research in small area estimates applications.

\section{Data}\label{sec-data}
\begin{figure}[H]
 \centering
 	\includegraphics[width=0.8\textwidth]{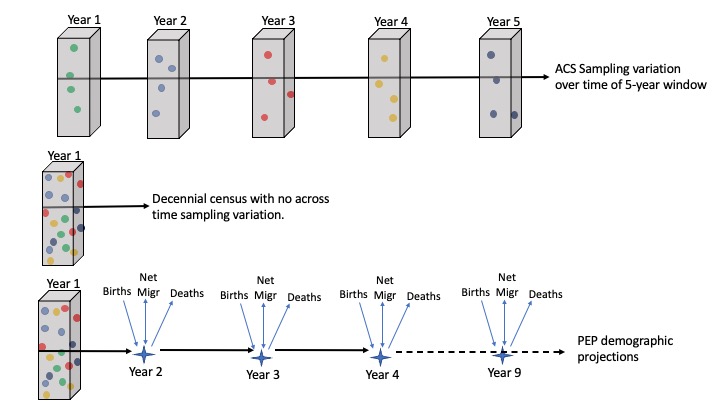}
 	\caption{Illustration of data generating processes for ACS, Decennial census, and PEP demographic projections. Boxes refer to data collection on individuals via surveys. Crosses refer to data generation through demographic projection formula. Color dots refers to blocks of individuals.}
 	\label{fig:loafofbread0}
 \end{figure}

\noindent
County-specific population size estimates typically come from the United States Census Bureau (USCB) in the form of (1) decennial census counts, (2) American Community Survey data, and (3) Population Estimation Program (PEP) reported estimates. Below we give brief descriptions of the different data sources and corresponding data collection methods. Figure \ref{fig:loafofbread0} illustrates the data generating process across the different UCSB programs. 

\subsection{Decennial Census}
The decennial census is a cross-sectional comprehensive survey constitutionally mandated to count every person in the U.S., which is accomplished through multiple modes of collection including mail, internet, phone-surveys, as well as door-to-door recruitment. Figure \ref{fig:loafofbread0} illustrates the data generating process in which the USCB seeks to count the entire population (colored dots) in the decennial year. The decennial census is considered the gold standard because it is comprehensive, and aims to capture the entire population. Due to complete enumeration, decennial counts do not suffer from sampling variation, thus can be used as a control for population sampling estimates in alternative surveys, e.g., ACS.  \\

The largest disadvantage of the decennial census is that data are only collected every 10 years, and often an additional two years are needed for tabulation. Rapidly changing areas and demographic or economic changes are not captured sufficiently quickly, particularly for public health surveillance applications. In addition, completion of the decennial census comes at enormous expense to ensure every household is sampled. Lastly, the decennial census population estimates may suffer from non-sampling errors, which can be attributed to under/over counts of the population (Spielman, S. et al. (2014); United States Census Bureau (2012)). \\

In an attempt to assess the extent of the non-sampling error affecting decennial census data, both nationally, and within particular sub-populations, the Census Coverage Measurement Program (CCM) conducts a post-enumeration survey following each decennial census, in which it measures the extent of erroneous enumerations due to duplications or other errors, incorrect geo-recorded residences, and omissions. The CCM program reports net undercount (coverage) measures, which capture the extent of erroneous enumerations and omissions (U.S. Census Bureau
: CCM (2012)). 
\begin{align}\label{eq:netundercount}
\text{Net Undercount} &= \text{truth} - \text{census}\\
\text{\% Net Undercount} & = \frac{truth - census}{truth} \cdot 100
\end{align}
\noindent
in which the estimated ``truth" is derived using the post-enumeration survey, and a dual system estimation method (DSE). The CCM reported, for Georgia specifically, the estimated 2010 census state count was 9434.5 (thousands), and the estimated percent state-level net undercount was $0.91\% (sd =1.04\%)$. The net undercount provides some information about the extent of non-sampling errors, and therefore, we incorporate the percent net undercount in our modeling of decennial census county-specific population counts (U.S. Census Bureau: CCM (2012)).

\subsection{American Community Survey data}
The ACS is a complex sample survey conducted by the U.S. Census Bureau annually. ACS population data are available in 1-year period estimates for all counties with population size $>65,000$, and in 5-year rolling averages for all census geographies. Figure \ref{fig:loafofbread0} illustrates that for the 5-year period estimates, ACS collects 3.5 million independent samples of data nationally (approximately 2.5\% of the population) for each year within the 5-year time interval (different color dots per year refer to different annual samples of the population), and aggregates weight-adjusted data to derive period estimates. ACS individual level data are aggregated to population level data using individual weights derived from sampling probabilities (US Census Bureau (2014)). Subsequently, post-stratification adjustments are made to ensure a degree of consistency between PEP reported population estimates (described in Section \ref{sec:pepi}), and ACS reported population estimates for the same corresponding county population. An important distinction between ACS and other USCB data sources is the time interval corresponding to its data collection. Although the data collection process spans 5-year intervals, ACS 5-year estimates are published annually, representing estimates from overlapping periods of data, i.e., 2010-2014, 2011-2015, etc. Therefore, these annually reported 5-year estimates are not independent. USCB recommends users refer to ACS 5-year estimates using the end-year and not the mid-year of the period (US Census Bureau (2018)). In addition to point estimates, ACS reports associated margins of error, which quantify the uncertainty (variability) due to sampling error. An advantage of the ACS relative to the decennial census is the availability of small area population estimates published annually. However, compared to the decennial census,  ACS estimates include sampling errors due to the small samples collected for small areas, and long data collection periods, with rotating sampling schemes across counties. ACS estimates may also be inconsistent with counts reported by the decennial census. USCB explicitly discourages the use of ACS 1- or 5-year population size estimates as denominator data, and instead recommends the use of PEP reported annual population estimates (US Census Bureau 2009;2018). However, it is common practice, in the literature, to use ACS 5-year estimates in small area analysis, particularly, in the absence of available PEP data for census tracts, and block groups (US Census Bureau 2009; 20014; 2018).
 \bigskip

 \subsection{Intercensal Projections}\label{sec:pepi}
The Population Estimation Program (PEP) produces annual intercensal population estimates, and projections, based on births, deaths, and migrations, for national, state, county, and micropolitan areas.  Population projections are derived using a cohort component model given below,\\
\begin{equation}\label{eq:demog}
\mbox{Population Base (y0) } + \mbox{ Births } - \mbox{ Deaths } + \mbox{ Net Migration } = \mbox{ Population Estimate (y1) }
\end{equation}
which assumes a deterministic relationship between the population rate of change, and the number of births, deaths, and migrations (Preston, S. et al. (2000)). 
The population base (starting population) is the last decennial census e.g., 2010  number reported, or the previous point in the time series. Figure \ref{fig:loafofbread0} illustrates that from the starting population, estimates are projected forward using births, deaths, and net migration estimates, and as such these estimates do not suffer from sampling error. The National Center for Health Statistics (NCHS) provides the vital statistics data used in these calculations. PEP uses a ``top-down" approach, in which national levels of population change are estimated by month, and by age, sex, race, and Hispanic origin, using Eq \ref{eq:demog}. They then produce estimates of the total annual population of counties, which sum to the state level. It is, generally, considered more reliable to estimate change for larger populations than aggregation upward from smaller population sizes (Population Estimation Program (2019)). First, month-specific population sizes at the national level by age, sex, race, and Hispanic ethnicity, are calculated. Then estimates of the total annual populations of counties sum to the state level. With national characteristics, state total, and county total estimates, the PEP program produces state, and county-level, estimates by age, sex, race, and Hispanic ethnicity. All estimates are consistent across geography and demographic characteristics, i.e., smaller areas sum to the larger areas. This requires an adjustment to final estimates to ensure consistency. While important, we defer incorporating uncertainty in vital statistics data to future work (Population Estimation Program (2019)). 
\bigskip

\subsection{Data Availability}\label{sec:dat}
A summary of data availability by data source is shown in Table \ref{tab:data}. Decennial population counts are reported for all geographical, and statistical areas, for every decennial year. ACS annual estimates are available for selected counties with populations greater than 65,000 (a little over 25\% of all counties), for years 2009-2018. ACS 5-year estimates are available for all geographical and statistical areas reported by census down to census tracts, and block groups.  Lastly, PEP projected annual estimates are available only for states, counties, and metro/micropolitan statistical areas, i.e., cities and towns, for 2010-2019. Information from the 2020 decennial census was not included due to the 2-year delay in reporting. 
\begin{table}[H]
\resizebox{\textwidth}{!}{
\begin{tabular}{ |p{3cm}||p{5cm}|p{3cm}|p{3cm}| }
 \hline
 \multicolumn{3}{|c|}{\textbf{Data availability by source} }\\
 \hline
\textbf{Data Source} & \textbf{Populations Available} & \textbf{Years Available} & \textbf{Related Sources} \\
 \hline
 Decennial census total counts &  All geographic areas, i.e., counties, census tracts, blocks  by race-stratified population & 2010 &  \\
 \hline
 ACS 5-year period sampled counts & All geographic areas, i.e., counties, census tracts, blocks by race-stratified population  & 2009-2018 by 5-year periods & Controlled with PEP reported counts  \\
 \hline
 PEP annual estimated counts   & All metropolitan and micropolitan statistical areas, counties, cities, towns by race-stratified population & 2010-2019 annual estimates & Decennial census used as base population \\
 \hline
\end{tabular}}
\caption{Data availability by USCB data source (decennial census, ACS,  and PEP projections). Availability is broken down by population in which counts are available, and the respective years for which data area reported. The last column indicates the dependencies between different data sources.}

\label{tab:data}
\end{table}

\begin{figure}[H]
\center
\includegraphics[width=0.8\textwidth, angle=0]{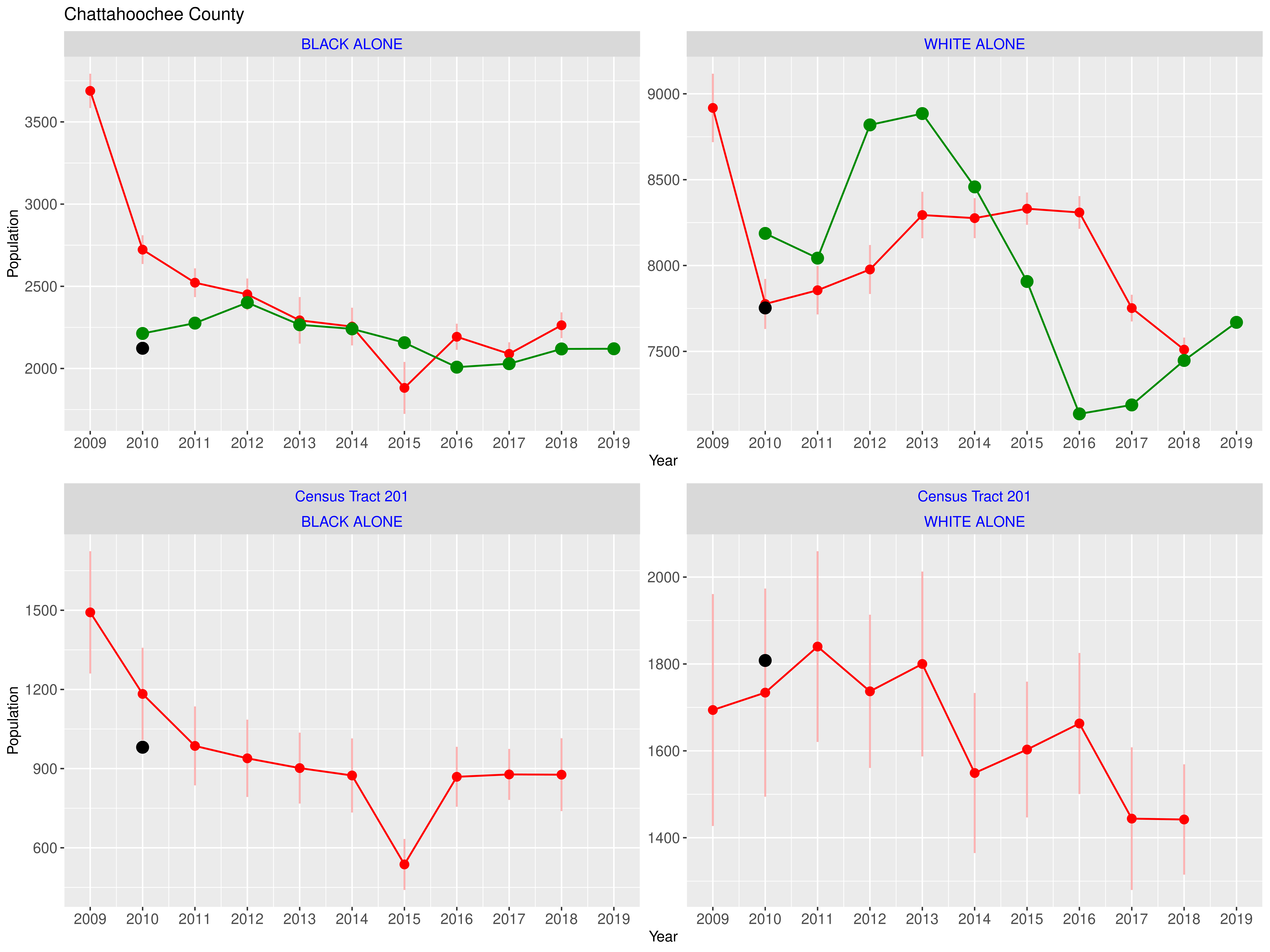}
\caption{Comparisons of ACS, decennial, and PEP reported county population counts for Chattahoochee county. Red refers to ACS reported counts, black refers to decennial census counts, and green refers to PEP projection estimates. Data shown for years 2009-2018 refer to the end years of ACS 5-year period estimates. We show data availability between county-specific data (top row), and census-tract level data (row 2).}
\label{fig:dat}
\end{figure}

As an illustration, Figure \ref{fig:dat} illustrates race-stratified population size estimates from each data source for Chattahoochee County, GA, for years 2009-2019. ACS population size estimates are shown in red, with decennial counts in black, and PEP estimates in green. We show non-trivial standard errors based on the reported ACS margins of error (vertical bars) assuming an approximate normal distribution, and using the transformation formula $SE = MOE/1.645$. Race-stratified ACS 5-year estimates, and associated errors, are available for both counties, and census-tracts within counties, illustrated using one example tract within Chattahoochee County (row 2). Decennial counts are available for all sub-groups down to tracts and block groups. Race-stratified PEP estimates are not available at the census tract level.\\

\bigskip

\section{Methods}\label{sec:mod}

\subsection{Summary:}
Our Bayesian Hierarchical Population Model (B-Pop) aims to integrate the three population data sources, and account for data source specific errors, in order to learn about the true population sizes. Figure \ref{fig:dag} shows a graphical representation of the Bayesian hierarchical model set-up. The main features of the model are  as follows:\\

(i) Estimates of county-race specific non-sampling errors, $\sigma_{NS[c,r]}^2$ (shown in the orange box), are informed by decennial census, PEP, and ACS population counts, as well as ACS reported design-based sampling error $s_{c,t,r}^2$ for county $c$, year $t$, race $r$, which is discussed in Section \ref{sec:vars}.  These unobserved non-sampling errors capture the uncertainty associated with decennial counts. Due to the complex relation between the three data sources, illustrated in Figure \ref{fig:dag}, we are able to use information from all three sources to inform non-sampling error estimates. As such, the variance, associated with each data source, can be written as a function of 
$s_{c,t,r}^2$.\\

(ii) The data model (defining the likelihood function) consists of modeling: (1) overlapping ACS 5-year weighted period estimates, (2) decennial census survey data,  and respective non-sampling error bias, and (3) PEP population rates of change.  The observed data quantities (shown in Figure \ref{fig:dag} in the blue boxes) are modeled using our specified likelihood functions. The data models define how we learn about the parameters of interest given the observed data quantities. The resulting likelihood functions are described in Section \ref{sec:datamod}.\\


(iii) True (but unobserved) log-transformed population counts are modeled for each for county $c$, year $t$, race $r$, $\eta_{c,t,r}$, using a random walk order 2 process model. To simplify development, no spatial correlation is assumed across counties. This process is described in detail in Section \ref{sec:proc}. Figure \ref{fig:dag} shows that the county-year-race stratified log-transformed true population estimates $\eta_{c,t,r}$ (shown in the red box) are informed by the hierarchical structures (shown in the orange box), as well as global hyper-parameters (shown in green). We learn about the true population given observed population data reported by ACS, PEP, and decennial census, respectively.\\
\bigskip

\noindent
Let $\bm{n}^{()}$ denote the vector of reported population counts for county $c$, race $r$, year $t$, in which we distinguish the data source within the superscript (), i.e. $\bm{n}^{(census)}$ denotes vectors of decennial census reported population counts. We denote corresponding vectors of $[c,t,r]$ true population estimates using $\mathbf{\gamma}$, and associated county-race specific non-sampling errors as $\bm{\sigma}^2_{NS}$. We remove subscripts of $[c,r,t]$, for the above quantities, for ease of readability. Let $\Theta$ be a vector of all hierarchical parameters, and hyper-parameters, shown in the orange box, illustrated in Figure \ref{fig:dag}. In Eqs. \ref{eq:full} to \ref{eq:cenpt} we show the full conditional posterior distribution to illustrate data dependencies. The posterior distribution of true population estimates is proportional to the joint data likelihood multiplied by prior assumptions, i.e.,
\begin{align}\label{eq:full} 
  p(\bm{\gamma}, \bm{\sigma}_{NS}^2, \Theta|\bm{n}^{(census)}, \bm{n}^{(PEP)}, \bm{n}^{(ACS)}, \bm{s}^2) \propto &
  p(\bm{n}^{(census)}, \bm{n}^{(PEP)}, \bm{n}^{(ACS)}|\bm{\gamma},  \bm{\sigma}_{NS}^2, \bm{s}^2, \Theta)  \quad \text{joint likelihood}\\ 
  &\cdot \quad p(\bm{\gamma}| \bm{\sigma}_{NS}^2, \Theta) \cdot p(\bm{\sigma}_{NS}^2, \Theta) \quad \text{priors} \nonumber
  \end{align}

Due to data dependencies, described in Section \ref{sec-data}, and illustrated in Figure \ref{fig:dag}, the joint likelihood can be broken down further into conditional likelihoods, shown in Eq. \ref{eq:acspt} to \ref{eq:cenpt}, in which the likelihood for ACS population counts is dependent upon PEP and census data, and similarly, the likelihood for PEP population counts is also dependent upon census data.  The hierarchical process model on $\bm{\gamma}$ is determined by the prior assumptions placed on hierarchical parameters and hyper-parameters, shown in Eq. \ref{eq:cenpt}.
  \begin{eqnarray}
  p(\bm{\gamma}, \bm{\sigma}_{NS}^2, \Theta|\bm{n}^{(census)}, \bm{n}^{(PEP)}, \bm{n}^{(ACS)}, \bm{s}^2)&=& p(\bm{n}^{(ACS)}| \bm{s}^{2}, \bm{n}^{(PEP)}, \bm{n}^{(census)}, \bm{\gamma}, \bm{\sigma}_{NS}^2, \Theta )\label{eq:acspt}\\
  &&\cdot \quad p(\bm{n}^{(PEP)}| \bm{n}^{(census)},\bm{\gamma}, \bm{\sigma}_{NS}^2, \Theta)  \label{eq:peppt} \nonumber\\
  && \cdot\quad  p(\bm{n}^{(census)}| \bm{\gamma}, \bm{\sigma}_{NS}^2, \Theta) \nonumber \\
  &&  \cdot \quad p(\bm{\gamma}| \bm{\sigma}_{NS}^2, \Theta) \cdot p(\bm{\sigma}_{NS}^2, \Theta)   \label{eq:cenpt} 
\end{eqnarray}

\begin{center}
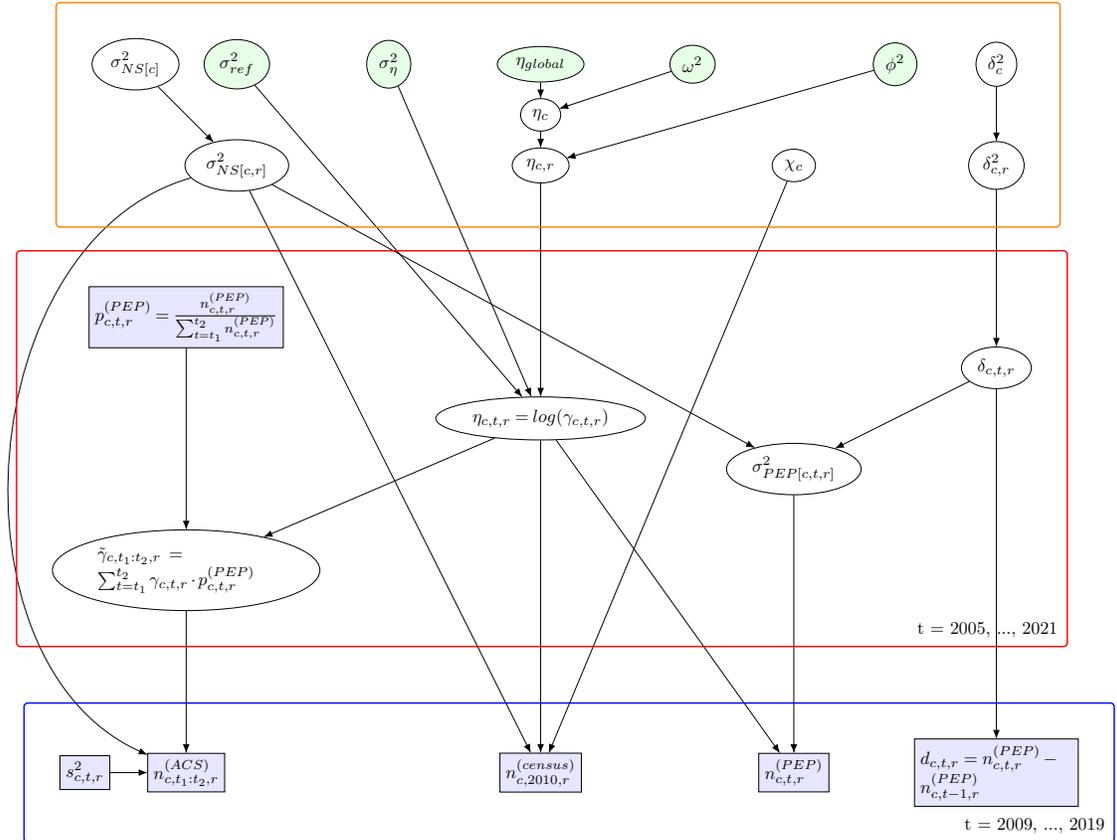
\begin{figure}[H]
    \resizebox{0.95\textwidth}{!}{%
\begin{tikzpicture}
\node[state, fill=green!10] (siget) at (2,14) {$\sigma^2_{\eta}$};
\node[state, fill=green!10] (phi) at (12,14) {$\phi^2$};
\node[state, fill=green!10] (omega) at (8,14) {$\omega^2$};
\node[state, fill=green!10] (etaglobal) at (5,14) {$\eta_{global}$};
\node[state, fill=green!10] (sigref) at (-1,14) {$\sigma^2_{ref}$};
\node[state] (deltac) at (14,14) {$\delta^2_{c}$};
\node[state] (deltacr) at (14,12) {$\delta^2_{c,r}$};
\node[state] (sigc) at (-3,14) {$\sigma^2_{NS[c]}$};
\node[state] (etac) at (5,13) {$\eta_{c}$};
\node[state] (xi) at (10,12) {$\chi_c$};
 \node[state] (signs) at (-1,12) {$\sigma^2_{NS[c,r]}$};
 \node[state] (etacr) at (5,12) {$\eta_{c,r}$};
 \node[state, rectangle,fill=blue!10] (p_pep) at (-2,9) {$p_{c,t,r}^{(PEP)} = \frac{n_{c,t,r}^{(PEP)}}{\sum_{t=t_1}^{t_2}n_{c,t,r}^{(PEP)}}$};
 \node[state] (eta) at (5,7) {$\eta_{c,t,r} = log(\gamma_{c,t,r})$};
  \node[state] (deltactr) at (14,8) {$\delta_{c,t,r}$};
   \node[state] (sigpep) at (10,6) {$\sigma^2_{PEP[c,t,r]}$};
    \node[state,text width=3.5cm] (gamtilde) at (-2,4) {$\tilde{\gamma}_{c,t_1:t_2,r} = \sum_{t=t_1}^{t_2} \gamma_{c,t,r} \cdot p_{c,t,r}^{(PEP)}$};
    \node[state, rectangle,fill=blue!10] (acs) at (-2,0) {$n^{(ACS)}_{c,t_1:t_2,r}$};
    \node[state,rectangle,fill=blue!10] (dec) at (5,0) {$n_{c,2010,r}^{(census)}$};
     \node[state,rectangle,fill=blue!10] (pep) at (10,0) {$n_{c,t,r}^{(PEP)}$};
     \node[state, rectangle, text width=3cm,fill=blue!10] (dctr) at (14,0) {$d_{c,t,r} = n_{c,t,r}^{(PEP)} - n_{c,t-1,r}^{(PEP)}$};
     \node[state, rectangle,fill=blue!10] (s) at (-4,0) {$s_{c,t,r}^{2}$};

\path (sigc) edge (signs);
\path (etaglobal) edge (etac);
\path (etac) edge (etacr);
\path (sigref) edge (eta);
\path (siget) edge (eta);
\path (phi) edge (etacr);
\path (omega) edge (etac);
\path (deltac) edge (deltacr);
 \path (p_pep) edge (gamtilde);
\path(signs) edge[bend left =-70] (acs);
    \path (etacr) edge (eta);
    \path (signs) edge (dec);
    \path (signs) edge (sigpep);
    \path (deltacr) edge (deltactr);
    \path (deltactr) edge (sigpep);
    \path (deltactr) edge (dctr);
    \path (sigpep) edge (pep);
     \path (s) edge (acs);
    \path (gamtilde) edge (acs);
     \path (eta) edge (dec);
      \path (eta) edge (gamtilde);
       \path (xi) edge (dec);
       \path (eta) edge (pep);

    \node[plate={t = 2009, ..., 2019}, blue,inner sep=20pt, fit=(acs) (dec) (pep) (s) (dctr)] (plate1) {};
     \node[plate={t = 2005, ..., 2021}, red,inner sep=20pt, fit=  (eta) (gamtilde) (deltactr) (p_pep)] (plate1) {};
      \node[plate={}, orange,inner sep=20pt, fit=(xi) (sigc) (etacr) (etac) (deltac) (deltacr) (signs) ] (plate1) {};

\end{tikzpicture}
}
\hspace{0.2cm}

\caption{
Directed graphical representation of the B-Pop hierarchical model. Blue rectangles denote observed data quantities, and circles denote latent variables (green shaded circles for global hyper-parameters). Solid arrows denote stochastic dependency.  Boxes group quantities by indices, i.e., (1) blue box contains observed population data, stratified by county-year-race for years 2009 to 2019, (2) red box contains both estimated parameters and observed data quantities stratified by county-year-race, for years 2005 to 2021, (3) orange box contains county-race,  county-specific, and global parameters. Subscripts refer to county $c$, year $t$, race $r$.}
\label{fig:dag}
\end{figure}
\end{center}

\subsection{Description of sources of data variability}\label{sec:vars}
We begin by listing sources of sampling, and non-sampling variability for each of the three data sources.
\paragraph{Decennial Error} In county $c$, and race group $r$, we assume that for all years the population counts have a time-invariant county-race specific non-sampling random error, denoted $\sigma^2_{NS[c,r]}$. Therefore, we assume each county-race combination to have a unique level of systematic non-random error, which is unmeasured, and does not vary across years. The non-sampling error, associated with decennial data, captures the systematic variability in decennial counts attributed to outside factors such as, but not limited to, non-response bias, and under-reporting. There is no information directly available to inform estimates of this systematic error, therefore, we apply a model-based approach, in which we use a Bayesian hierarchical model to estimate $\sigma^2_{NS[c,r]}$. We hierarchically model county-race stratified non-sampling errors, which are drawn from a truncated-normal distribution centered around county-specific non-sampling errors, shown in Eq. \ref{eq:signns}: \\
 \begin{align}\label{eq:signns}
 \sigma^2_{NS[c,r]} & \sim N_{[0,]}(\sigma^2_{NS[c]}, 100)\\
\sigma^2_{NS[c]} & \sim N_{[0,]}(0, 100)\nonumber
\end{align}
\noindent

\paragraph{ACS Error} Outside of the decennial year, we allow for additional variability in population size estimates attributable to the ACS sampling design. We denote this sampling variability $s^2_{c,t,r}$, and treat these values as known constants derived from ACS reported margins of error (MOEs),  $s^2_{c,t,r} = MOE_{c,t,r}/1.645$. In summary, for county-race $[c,r]$, in year $t=2010$, ACS associated variance is equal to non-sampling error alone, and in years $t\neq 2010$, the  variability is given by $\sigma_{NS[c,r]}^2 + s_{c,t,r}^2$.  

\begin{align}\label{eq:acserrr}
\sigma^2_{c,t,r} &=\begin{cases}
   \sigma^2_{NS[c,r]},& \text{if } t = 2010\\
 \sigma^2_{NS[c,r]} + s^2_{c,t,r},  & \text{otherwise}
    \end{cases}
\end{align}

\paragraph{PEP Error} In the decennial year, PEP associated variance is equal to non-sampling error alone. Outside the decennial year, there is additional variability, associated with PEP counts, due to uncertainty in the population deviation term (the annual population change). Importantly, the extent of uncertainty increases with increasing distance from the decennial year. As an example, the variance of 2011 PEP reported counts is equal to the variance of the decennial counts, plus the variance of the annual population change for 2011. In contrast, the variance of 2012 PEP reported counts is equal to that of 2011 PEP counts plus the variance associated with the 2012 annual population change. 
\begin{equation}
var\left(n_{c,t,r}^{(PEP)}\right) = var(n_{c,2010,r}^{(census)}) + \sum_{k=2010}^{t} var(d_{c,t,r})
\end{equation}
\noindent
Therefore, the total variance, associated with PEP counts, is the sum of non-sampling error and the variance of the annual population change, shown in Eq. \ref{eq:pperr}.
\begin{align}\label{eq:pperr}
\sigma^2_{PEP[c,t,r]} &=\begin{cases}
   \sigma^2_{NS[c,r]},& \text{if } t = 2010\\
 \sigma^2_{NS[c,r]} + \sum_{k=2010}^{t}\delta^2_{c,k,r},  & \text{otherwise}
    \end{cases}\\
   \delta_{c,t,r} & \sim N_{[0,]}(\delta_{c,r}, 100)\nonumber \\
   \delta_{c,r} & \sim N_{[0,]}( \delta_c, 100)\nonumber \\
 \delta_{c} & \sim N_{[0,]}( 0, 100)\nonumber
\end{align}
\bigskip

\subsection{Data Models}\label{sec:datamod}

The data model defines the assumed distribution for the observed data. It associates the vector of observed population counts for county $c$, year $t=2009,...,2019$, and race $r$, $n_{c,t,r}$, to the parameters of interest, the true population counts for corresponding county, year, and race, $\gamma_{c,t,r}$. We denote the data source for each vector of population counts with superscript (), i.e., $n_{c,r}^{(census)}$ refers to the 2010 census reported population count for county $c$, race $r$. We incorporate hierarchical data models for each of the different data collection and processing methodologies, in order to model the true population sizes given the data we have observed. Building off of the sources of variability for each data component, we consider a model of each component assuming the others are fixed.\\

\paragraph{Decennial census} In the likelihood component associated with the decennial census population counts for county $c$, and race $r$, $n_{c,r}^{(census)}$, we incorporate the estimated percent net undercount within the data model, to account for bias related to non-sampling errors, see Section 2.1. Let
$\gamma_{c,2010,r}$ denote the true population for county $c$, in decennial year 2010, race $r$. Eq. 2 shows how the percent net undercount (non-sampling error bias) relates the census count to the true population count. We assume the decennial population counts, given the true population counts, to be normally distributed, and write the expected mean as a function of the estimated county-level net percent undercount $\chi_c$, and the true population estimate $\gamma_{c,2010,r}$. We assume that the county-specific net coverage, $\chi_c$, does not deviate significantly from the CCM reported state-level percent net coverage of 0.91. Therefore, $\chi_c$ is modeled hierarchically, assuming a truncated normal distribution bounded below by 0, denoted $N_{[0,]}()$. The posterior distribution of census alone, assuming others are fixed, is given by,
\begin{align}\label{eq:postdec3}
    p(\bm{\gamma}_{2010}|  \bm{n}^{(census)} ,  \bm{\chi}, \bm{\sigma}^2_{NS},\bm{\Theta} )\propto &   p(\bm{n}^{(census)}| \bm{\gamma}_{2010},\bm{\chi}, \bm{\sigma}^2_{NS})\cdot &  \quad \text{  (likelihood})  \\ 
   &  p(\bm{\Theta})\cdot p(\bm{\chi})  \cdot     p( \bm{\sigma}^2_{NS}|\bm{\Theta}) \cdot   p( \bm{\gamma}_{2010}| \bm{\Theta}) \cdot  & \quad \text{(hierarchical priors)}  \nonumber\\
  =& \prod_c^C \prod_r^R N\left(n_{c,r}^{(census)} |\gamma_{c,2010,r}, \chi_c, \sigma^2_{NS[c,r]}\right)\cdot & \nonumber\\
&      p( \bm{\sigma}^2_{NS}|\bm{\Theta}) \cdot   p( \bm{\gamma}_{2010}| \bm{\Theta}) \cdot p(\bm{\chi})  \cdot  p(\bm{\Theta}) \nonumber \\\nonumber
\end{align}
\noindent
in which $\bm{n}^{(census)}$ refers to the vector of observed county-race stratified decennial census population counts, $\bm{\gamma}_{2010}$ refers to the vector of county-race stratified estimates of the true population count in 2010. Let $\chi_c$ refer to the net coverage parameter, and $\sigma^2_{NS}$ to the vector of errors associated with county-race stratified decennial counts. Lastly, $\bm{\Theta}$ refers to the combination of hyper-parameters, and parameters, which capture our prior assumptions about $\gamma_{c,2010,r}$ based on the hierarchical structure described further in Section \ref{sec:proc}. The overall likelihood function, the decennial census data model is given by, 
\begin{align}\label{eq:decdat}
    n_{c,r}^{(census)}|\gamma_{c,2010,r}, \chi_c, \sigma^2_{NS[c,r]} &\sim N\left(\gamma_{c,2010,r} \cdot \left(1 - \frac{\chi_c}{100}\right), \sigma^2_{NS[c,r]}\right)\\
    \chi_c &\sim N_{[0,]}(0.91, 1.04^2)\nonumber
\end{align}
\noindent
The error variance for the decennial counts is solely the estimated non-sampling errors, $\sigma^2_{NS[c,r]}$, which we model hierarchically, and is further described in our full model in Section \ref{sec:vars} above.
\bigskip

\paragraph{PEP data} PEP population counts for county $c$, year $t$, race $r$, $n_{c,t,r}^{(PEP)}$, capture annual true population estimates. The vector of county-year-race stratified estimates of the true population, $\bm{\gamma}$, are informed by corresponding vectors of county-year-race specific PEP reported population counts, $\bm{n}^{(PEP)}$, dependent on the census reported population size $\bm{n}^{(census)}$ plus annual population change $\bm{d}$. Errors, associated with PEP reported population counts $\bm{\sigma}^2_{PEP}$, are dependent on estimated non-sampling errors $\bm{\sigma}^2_{NS}$, and errors associated with annual population change, $\bm{\delta}^2$. Lastly, $\bm{\Theta}$ refers to the combination of hyper-parameters, and parameters, which capture our prior assumptions about $\gamma_{c,t,r}$, further described in Section \ref{sec:proc} below. The posterior distribution for PEP alone, given others are fixed, is given by,
\begin{align}\label{eq:postdec1}
    p(\bm{\gamma},\bm{\sigma}^2_{PEP}, \Theta |  \bm{n}^{(PEP)} ) &\propto  \underbrace{ p\left(\bm{n}^{(PEP)}| \bm{\gamma},\bm{\sigma}^2_{PEP}\right) } \quad & \text{  (likelihood}) \\ \nonumber
   &  \cdot \underbrace{p(\bm{\sigma}^2_{PEP}|\Theta) } \cdot p( \bm{\gamma}| \Theta) \cdot p(\Theta) &\text{ (hierarchical priors)}  \\ \nonumber
 & =  \textcolor{blue}{p\left(\bm{n}^{(census)} + \bm{d}| \bm{\gamma},\bm{\sigma}^2_{PEP}\right)}   & \text{  (Based on Eq. \ref{eq:pepp})}\nonumber\\ \nonumber
&  \cdot \textcolor{blue}{p(\bm{\sigma}^2_{NS} + \bm{\delta}^2|\Theta)}\cdot p( \bm{\gamma}| \Theta)\cdot p(\Theta) &\text{ (Based on Eq. \ref{eq:sigpep})}\\ \nonumber
\end{align}

This yields,
\begin{eqnarray}
  p(\bm{\gamma},\bm{\sigma}^2_{PEP}, \Theta |  \bm{n}^{(PEP)} ) &\propto  \prod_c^C \prod_r^R\prod_t^T N\left(n_{c,r}^{(census)} + \sum_{t} d_{c,t,r} |\gamma_{c,t,r}, \sigma^2_{NS[c,r]} , \delta^2_{c,t,r}\right)\\
 & \cdot  p(\bm{\sigma}^2_{NS} + \bm{\delta}^2|\Theta) \cdot  p( \bm{\gamma}| \Theta) \cdot p(\Theta) \nonumber
\end{eqnarray}


As such, we assume a data model structure in which these counts are normally distributed, centered around the true population estimate $\gamma_{c,t,r}$. 

\begin{align}
    n_{c,t, r}^{(PEP)}|n_{c,r}^{(census)}, \gamma_{c,t,r},\sigma_{PEP[c,t,r]}  &\sim N(\gamma_{c,t,r} , \sigma_{PEP[c,t,r]}^2)\\\label{eq:pepp}
     n^{(PEP)}_{c,t,r} & =  n^{(census)}_{c,r}  + \sum_{k=2011}^t d_{c,k,r}\\
     d_{c,t,r}& =n_{c,t,r}^{(PEP)} - n_{c,t-1,r}^{(PEP)}
\end{align}

We include a likelihood function which captures annual population deviations from one year to the next $(d_{c,t,r})$, or in other words, the net-growth of a population from time $t-1$ to time $t$, which is equal to the sum of births, deaths, and net migrations for year $t$, i.e., $d_{c,t,r} =n_{c,t,r}^{(PEP)} - n_{c,t-1,r}^{(PEP)} = \text{birth}_{c,t,r} + \text{deaths}_{c,t,r} + \text{net migration}_{c,t,r}$. The annual population changes are modeled assuming a normal distribution, centered around 0, with the unknown variance associated with population deviations $\delta_{c,t,r}^2$, which is modeled hierarchically, and is further detailed in Eq. \ref{eq:pperr} (above). Therefore, $\delta_{c,t,r}$ represents the variance associated with a one year change in the population. 
\begin{align}
    d_{c,t,r} &= n_{c,t,r}^{(PEP)} - n_{c,t-1,r}^{(PEP)}\\
    d_{c,t,r}| \delta_{c,t,r}  & \sim N(0, \delta_{c,t,r}^2)
\end{align}

In the decennial year, the deviation term $d_{c,t,r}$ is equal to zero, and the total error variance associated with PEP counts, $\sigma_{PEP[c,t,r]}^2$, is equal to non-sampling error alone $\sigma^2_{NS[c,r]}$. Outside the decennial year, the error variance is equal to non-sampling error plus the cumulative sum annual change variances, $\delta_{c,t,r}$, shown in Eq. \ref{eq:sigpep}.  For simplicity of development, we assume independence between reported counts, which is not a realistic assumption, therefore, potential future work may focus on incorporating covariance metrics in the sum of the variances.

\begin{equation}\label{eq:sigpep}
    \sigma_{PEP[c,t,r]}^2 = \begin{cases}
    \sigma^2_{NS[c,r]},&\text{if } t = 2010\\
     \sum_{k=2010}^{t}\delta^2_{c,r,k} + \sigma^2_{NS[c,r]},& \text{if } t \neq 2010
\end{cases}\\
\end{equation}

As an example, for the year 2012, the PEP reported count is equal to the census count plus the annual population changes for years 2011 and 2012, shown in Eq. \ref{eq:ex2012}. Assuming independence of demographic changes each year, the variance associated with PEP reported counts for county $c$, race $r$, in 2012 is derived below,

\begin{align}\label{eq:ex2012}
   n^{(PEP)}_{c,2012,r} & =  n^{(PEP)}_{c,2012,r}  + d_{c,2012,r}\\
    n^{(PEP)}_{c,2012,r} & =  n^{(census)}_{c,r}  + d_{c,2011,r} + d_{c,2012,r}\\
     \sigma_{PEP[c,2012,r]}^2 &=   var(n^{(census)}_{c,r})  + var(d_{c,2011,r}) + var(d_{c,2012,r})\\
     &= \sigma^2_{NS[c,r]} + 
     \sum_{k=2010}^{2012}\delta^2_{c,r,k}\nonumber
     \end{align}

\paragraph{ACS data} We denote the vector of observed 5-year population estimates for all years in the study period for county $c$, and race $r$, by  $\bm{n}^{(ACS)}_{c,r}$,  and assume it follows a multivariate normal distribution, given the true population counts, to account for the correlated data structure. To relate the 5-year ACS rolling averages to the true annual population counts, we introduce $\tilde{\gamma}_{c,t_1:t_2,r}$, the true 5-year weighted average, from start year $t_1$ to end year $t_2$, of the annual county-year-race specific true counts $\gamma_{c,t,r}$, with county-race-year specific weights denoted by $p^{(PEP)}_{c,t,r}$. We define $p^{(PEP)}_{c,t,r}$ as the observed PEP proportion of the annual population in year $t$ out of the total aggregated population over the 5-year period, i.e., $p^{(PEP)}_{c,t,r} = \frac{n_{c,t,r}^{(PEP)}}{\sum_{t = t1}^{t2} n_{c,t,r}^{(PEP)}}, \text{ for } t = t_1,..., t_2 = t_1 + 5$.  In summary, we use the annual county-year-race proportion breakdowns, reported by PEP data, to weight the annual parameter estimates across the 5-year period. We use $\bm{n}^{(ACS)}$ to refer to vector of county-race stratified ACS population period estimates, with corresponding vectors of 5-year weighted averages $\bm{\tilde{\gamma}}$. The variance-covariance matrix $\bm{\Sigma}$ consists of non-sampling error $\bm{\sigma}^2_{NS}$, and observed ACS reported standard errors $\bm{s}^2$. Let $\bm{\Theta}$ refers to the combination of hyper-parameters and parameters, which capture prior assumptions, further detailed in Section \ref{sec:proc}. The posterior distribution is given by,
\begin{align}\label{eq:postdec2}
    p(\bm{\tilde{\gamma}},\bm{\Sigma}, \Theta |  \bm{n}^{(ACS)} ) &\propto   p\left(\bm{n}^{(ACS)}| \bm{\tilde{\gamma}},\bm{\Sigma}\right) & \text{  (likelihood}) \\
     & \cdot  \underbrace{p(\bm{\Sigma}|\Theta) } \cdot \underbrace{p( \bm{\tilde{\gamma}}| \Theta)} \cdot p(\Theta)&  \text{(hierarchical priors)}  \nonumber\\ 
&  =  p\left(\bm{n}^{(ACS)}| \bm{\tilde{\gamma}},\bm{p}^{(pep)}, \bm{\sigma}^2_{NS}, \bm{s}^2\right)\nonumber\\ 
&\cdot \textcolor{blue}{p(\bm{\sigma}^2_{NS} + \bm{s}^2|\Theta)} \cdot \textcolor{blue}{p(\bm{\gamma} \cdot \bm{p}^{(pep)}|\Theta)} \cdotp(\Theta)  &  \text{(Based on Eqs. \ref{eq:avgs}- \ref{eq:acserr})}\nonumber
\end{align}
This yields,
\begin{align}
  p(\bm{\tilde{\gamma}},\bm{\Sigma}, \Theta |  \bm{n}^{(ACS)} ) &\propto  \prod_c^C \prod_r^R\prod_t^T N\left(n_{c,r}^{(ACS)}  |\gamma_{c,t,r}, p_{c,t,r}^{(pep)}, \sigma^2_{NS[c,r]} , s^2_{c,t,r}\right)\\
 & \cdot p(\bm{\sigma}^2_{NS} + \bm{s}^2|\Theta) \cdot  p( \bm{\gamma}\cdot \bm{p}^{(pep)}| \Theta)\cdot p(\Theta) \nonumber
\end{align}

We define the ACS data likelihood as follows:\\
\begin{align}
\bm{n}^{(ACS)}_{c,r}| \bm{n}^{(PEP)}_{c,2005:2018,r}, \bm{\tilde{\gamma}}_{c,r} & \sim N_T(\bm{\tilde{\gamma}}_{c,r}, \Sigma_{c,r,1:T,1:T})
\end{align}
in which $\bm{n}^{(ACS)}_{c,r} = \left(n^{(ACS)}_{c,2009,r}, n^{(ACS)}_{c,2010,r}, ..., n^{(ACS)}_{c,2018,r}\right)$ denotes the vector of ACS reported population counts for county $c$, race $r$, and the corresponding vector of true 5-year weighted averages denoted by, 
\begin{align}
\bm{\tilde{\gamma}}_{c,r} &= \left(\tilde{\gamma}_{c,2005:2009,r}, \tilde{\gamma}_{c,2006:2010,r}, ..., \tilde{\gamma}_{c,2013:2018,r}\right)\nonumber\\ \label{eq:avgs}
\tilde{\gamma}_{c,t1:t2,r} & = \left(\sum_{t = t1}^{t2} \gamma_{c,t,r}\cdot p_{c,t,r}^{(PEP)}\right)\\
\gamma_{c,t,r} &= exp(\eta_{c,t,r})\nonumber
\end{align}
The variance-covariance matrix is defined by:
\begin{align}
\Sigma_{c,r,1:T,1:T} & = \begin{bmatrix}
\sigma^2_{c,r,2009} & \rho \cdot \sigma_{c,r,2009}\sigma_{c,r,2010} & \ldots &  \rho \cdot \sigma_{c,r,2009}\sigma_{c,r,2018}\\
\rho \cdot \sigma_{c,r,2010}\sigma_{c,r,2009} & \sigma^2_{c,r,2010} & \ldots& \rho \cdot \sigma_{c,r,2010}\sigma_{c,r,2018}\\
\vdots & \vdots & \ddots & \vdots \\
\rho \cdot \sigma_{c,r,2018}\sigma_{c,r,2009} & \ldots & \ldots &  \sigma^2_{c,r,2018}\\
\end{bmatrix}\\
& \nonumber \\ \label{eq:acserr}
\end{align}
We define $\sigma_{c,t,r}^2$ as given in Eq. \ref{eq:acserrr}, and shown below:
\begin{align*}
\sigma^2_{c,t,r}&= \begin{cases}
   s^2_{c,t,r} + \sigma^2_{NS[c,r]},& \text{if } t \neq 2010\\
    \sigma^2_{NS[c,r]},&\text{if } t = 2010
\end{cases}
\end{align*}
\noindent
For ease of readability, let $\sigma_{c,t,r2}$ denote the variance term corresponding to county $c$, race $r$, 5-year period $t_1:t_5$. The correlation parameter $\rho$ is modeled with a $Unif(0,1)$ prior. 
\bigskip

\subsection{Modeling unobserved log-transformed population counts over time}\label{sec:proc}
To estimate the true population counts for county $c$, year $t \in (2005, 2011, ..., 2021)$, and race group $r$, denoted by $\gamma_{c,t,r}$, we propose a hierarchical random-walk model of second-order differences, on the log-transformed true population counts, $\eta_{c,t,r}=\text{log}(\gamma_{c,t,r})$. Modeling log-transformed population counts allows a flexible way to share information across counties using a hierarchical structure, and to predict estimates for years without data. Notably, for the sake of simplicity, we do not incorporate spatial correlation in estimation of small area true population counts. In future work, we intend to use B-Pop small area population estimates within a small area disease mapping framework, which includes spatially structured random effects. 

We anchor our model around the decennial year, i.e., $t_{ref} = 2010$. The true log-population size in non-reference year $t$, $\eta_{c,t,r}$, is assumed to be equal to $2\cdot\eta_{c,t-1,r}$  minus $\eta_{c,t-2,r}$, shown in Eq. \ref{eq:sumterm1}. We include two variance parameters within the random-walk because we assume that the uncertainty associated with the decennial year $\sigma_{ref}^2$ is smaller compared to that outside of the decennial year $\sigma^2_{\eta}$. We apply a hierarchical structure in the reference year by adding a county-race specific level for the log-transformed population count, centered around a county-level, and then a global distribution, shown in Eqs. \ref{eq:cr1}- \ref{eq:cce}.  Lastly, the start year of 2005 was chosen such that the first 5-year period comprises years 2005-2009, with the end year corresponding to the observation year, i.e., $\tilde{\gamma}_{c,2005:2009, r} = \sum_{t = 2005}^{2009} \gamma_{c,t,r} \cdot p_{c,t,r}^{(PEP)}$.

\begin{align}\label{eq:sumterm1}
\eta_{c,t,r} & \sim N(2\cdot\eta_{c,t-1,r} -\eta_{c,t-2,r}, \sigma_{\eta}^2), \text{ for } t \neq 2010 \quad \text{(RW(2) prior)} \\
\eta_{c,t_{ref},r} & \sim N(\eta_{c,r} , \sigma_{ref}^2), \text{ for } t = 2010 \\
\eta_{c,r} & \sim N(\eta_{c}, \phi^2)\label{eq:cr1}\\
\eta_c & \sim N(\eta_{global},  \omega^2)\\ 
\eta_{global} & \sim N(0, 10^2) \label{eq:cce}\\
\sigma_{ref}, \sigma_{\eta}, \phi, \omega &\sim N_{[0,]}(0, 100)\label{eq:vars1}
\end{align}
\noindent
Vague priors were used on the hyper-parameters, in which we assigned truncated normal distributions $N_{[0,]}(0,100)$ for all variance hyper-parameters, and a vague $N(0,10^2)$ for the global parameter.


\subsection{Derivation of tract specific 1-year estimates} 
As mentioned in Section \ref{sec:dat}, census tract counts are reported by USCB only in the form of ACS 5-year estimates and decennial cross-sectional counts. Therefore, one of our aims is to obtain 1-year tract-race level estimates of the true population to fill in this missing piece of data reporting.
To estimate true population counts for tract $g$ within county $c$, year $t$, and race group $r$ $(\eta_{c[g], t, r})$ for $g=1,...,G_c$, we apply ACS-derived proportions to the county-specific population estimates. As such, we are able to propagate uncertainty related to estimated population counts, but we do not incorporate uncertainties associated with the ACS reported proportions. The process is given below:\\
\begin{enumerate}
    \item Let $\gamma^{(s)}_{c,t,r}$ denotes posterior sample estimates $(s = 1,...,S)$ of true population for county $c$, year $t$, race $r$.
    \item Let $p^{(ACS)}_{c[g],t,r}$ denote the ratio of ACS reported population for tract $c[g]$, year $t$, race $r$ to the sum of the population across tracts within county $c$, i.e., tract-specific population proportions of the county totals.
    \begin{equation}
p^{(ACS)}_{c[g],t,r} = \frac{n_{c[g],t,r}^{(ACS)}}{\sum_{g=1}^{G_c} n_{c[g],t,r}^{(ACS)}}
    \end{equation}
    \item To obtain tract-year-race specific true population sample estimates, we multiply the true county-population sample estimates to the ACS tract-specific proportions:
        \begin{equation}
\gamma^{(s)}_{c[g],t,r} = \gamma^{(s)}_{c,t,r} \cdot p^{(ACS)}_{c[g],t,r} 
    \end{equation}
    \item ACS tract-specific proportions are available for years 2009-2018. To get ACS tract-specific estimates for 2019, 2020, and 2021, we apply the ACS proportion reported in the last available year (2018). Conversely, to derive tract estimates for years 2005-2008, we apply the ACS proportion reported in the earliest year (2009).
    \item Lastly we obtain median and $95\%$ quantile estimates of the tract-year-race specific true population posterior samples.
 
\end{enumerate}

\subsection{Model validation}\label{validmethods}\label{sec-val}
We carried out model validation to check the predictive performance of the B-Pop model population estimates relative to  both PEP, and ACS,  reported population counts. Model performance was assessed through the a validation exercise, in which we left out PEP, and ACS, reported data for years 2016-2019. We fit the B-Pop model to the training set, and obtained posterior population estimates for left-out years. We summarized the difference between the model-based predictions and the PEP, and ACS, reported population sizes in terms of errors, i.e., the difference between the log-transformed observed population estimate and B-Pop point estimate, and coverage of 95\% prediction intervals. %
The procedure is summarized in  Figure ~\ref{fig:validmeth1}.

\begin{figure}[H]
\fbox{
\begin{minipage}{38em}
\textbf{Calculation of outcome measures in the validation exercise}
\begin{enumerate}
\item Model fitting: Fit the B-Pop model to the training data and obtain posterior samples $\eta^{(s)}_{c,t,r}$  for years with left-out data in the test set (years $=$ 2016-2019).
\item Predict population size for county $c$, year $t$, race $r$ for left-out observation by calculating posterior median of samples $\eta^{(s)}_{c,t,r}$. We denote $c[i], t[i], r[i]$ to correspond to the county-year-race of left-out observation $i$.
\[\hat{\eta}_{c[i], t[i], r[i]} = median(\eta^{(s)}_{c[i],t[i],r[i]})\]
\item Error calculation: Calculate the difference between observed log-transformed population (for both ACS and PEP) and the median estimate from the predictive distribution:
   \[error^{(pep)}_i = \left(log(n)^{(pep)}_{c[i], t[i], r[i]} - \hat{\eta}_{c[i], t[i], r[i]}\right)\]
   \[error^{(acs)}_i = \left(log(n)^{(ACS)}_{c[i], t[i], r[i]} - \hat{\eta}_{c[i], t[i], r[i]}\right)\]
Various summaries of the errors are reported. 
\item Calibration: Calculate the proportion of observed PEP and observed ACS population counts above and below their respective 95\% prediction interval. 
\end{enumerate}
\end{minipage}
}
\caption{Overview of calculation of errors and coverage of prediction intervals in out-of-sample validation exercises.}\label{fig:validmeth1}
\end{figure}

\subsection{Computation}
\noindent
We extract ACS, decennial, and PEP reported population estimates, and ACS margins of error, for 159 counties in Georgia, years 2009-2018, using the tidycensus package (Walker, K. (2020)) For model processing and output, a Markov Chain Monte Carlo (MCMC) algorithm samples from the posterior distribution of the parameters via the software $JAGS$ (Plummer, M. (2017)). Eight parallel chains were run with a total of 80,000 iterations in each chain. Of these, the first 20,000 iterations in each chain is discarded. Standard diagnostic checks using traceplots were used to check convergence (Plummer, M. (2017)).

\section{Results}\label{sec:res}

\subsection{Global estimates}\label{globest}
Table \ref{tab:global} shows the median point estimates and associated $95\%$ credible intervals (CIs) for the different variance components as well as the global level of log-transformed population estimates resulting from the application of the B-Pop model to the GA county-level data.
Notably, the global level of log counts, across county, times, and race stratifications, was estimated to be 9.14 with a $95\%$ CI $= (9.01, 9.27)$, which corresponds to a global average of 9,286 individuals (8,141, 10,577) on the non-transformed scale. Posterior estimates of variance, and correlation components,  are given in Table \ref{tab:global}. To view plots of the posterior distribution of the variance hyper-parameters against their prior assumptions, refer to Appendix C.

\begin{table}[H]
\begin{center}
 \begin{tabular}{||c c ||}
 \hline
 Global Parameter & point estimate ($95\%$ CIs)  \\ [0.5ex]
 \hline\hline
 \hline
 $\eta_{global}$ & 9.14 (9.01, 9.27) \\
 \hline
 $\sigma_{ref}$ & 1.07 (0.13, 1.44) \\
 \hline
 $\sigma_{\eta}$ & 0.017 (0.01, 0.02) \\
 \hline
 $\sigma_{\delta}$ & 0.007 (0.003, 0.014) \\
 \hline
 $\phi$ & 0.86 (0.15, 1.36) \\
 \hline
 $\omega$ & 0.77 (0.56, 0.92)\\
 \hline
 $\rho$ & 0.52 (0.51, 0.53)  \\
 \hline
 \hline
\end{tabular}
 \caption{Posterior estimates of global variance parameters and the global level of log-transformed population counts.}
  \label{tab:global}
\end{center}
\end{table}

\subsection{County estimates of non-sampling errors}
\begin{figure}[H]
 \centering
 	\includegraphics[width=0.8\textwidth]{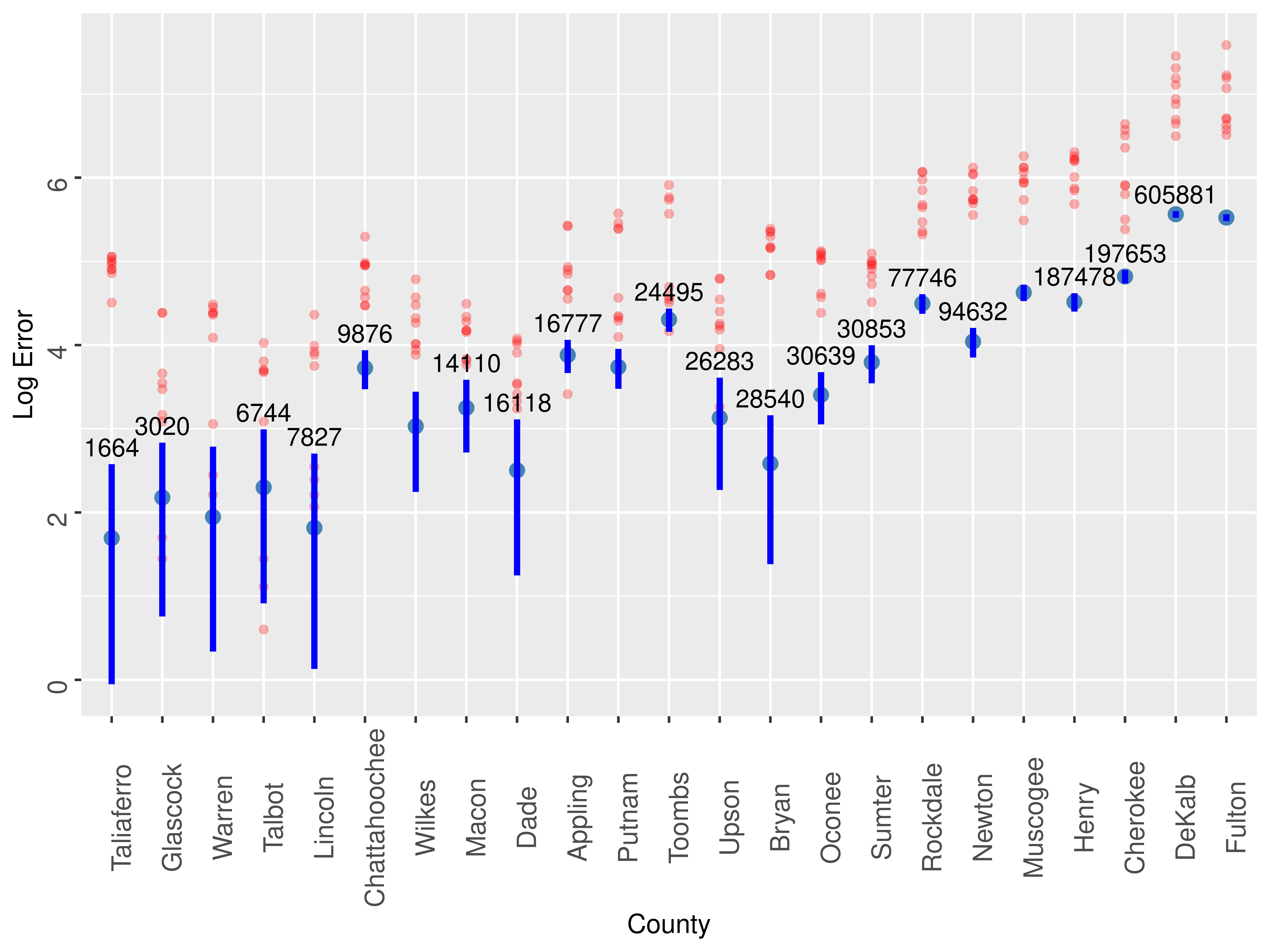}
 	\caption{County specific median estimates of log-transformed non-sampling error, and associated $95\%$ credible intervals shown in blue for selected counties (n=23) in Georgia. ACS reported sampling errors within county are shown in red. Counties are ordered by increasing population (2010 Census) shown in black.}
 	\label{fig:nonsamperr}
 \end{figure}

\begin{figure}[H]
 \centering
 	\includegraphics[width=0.8\textwidth]{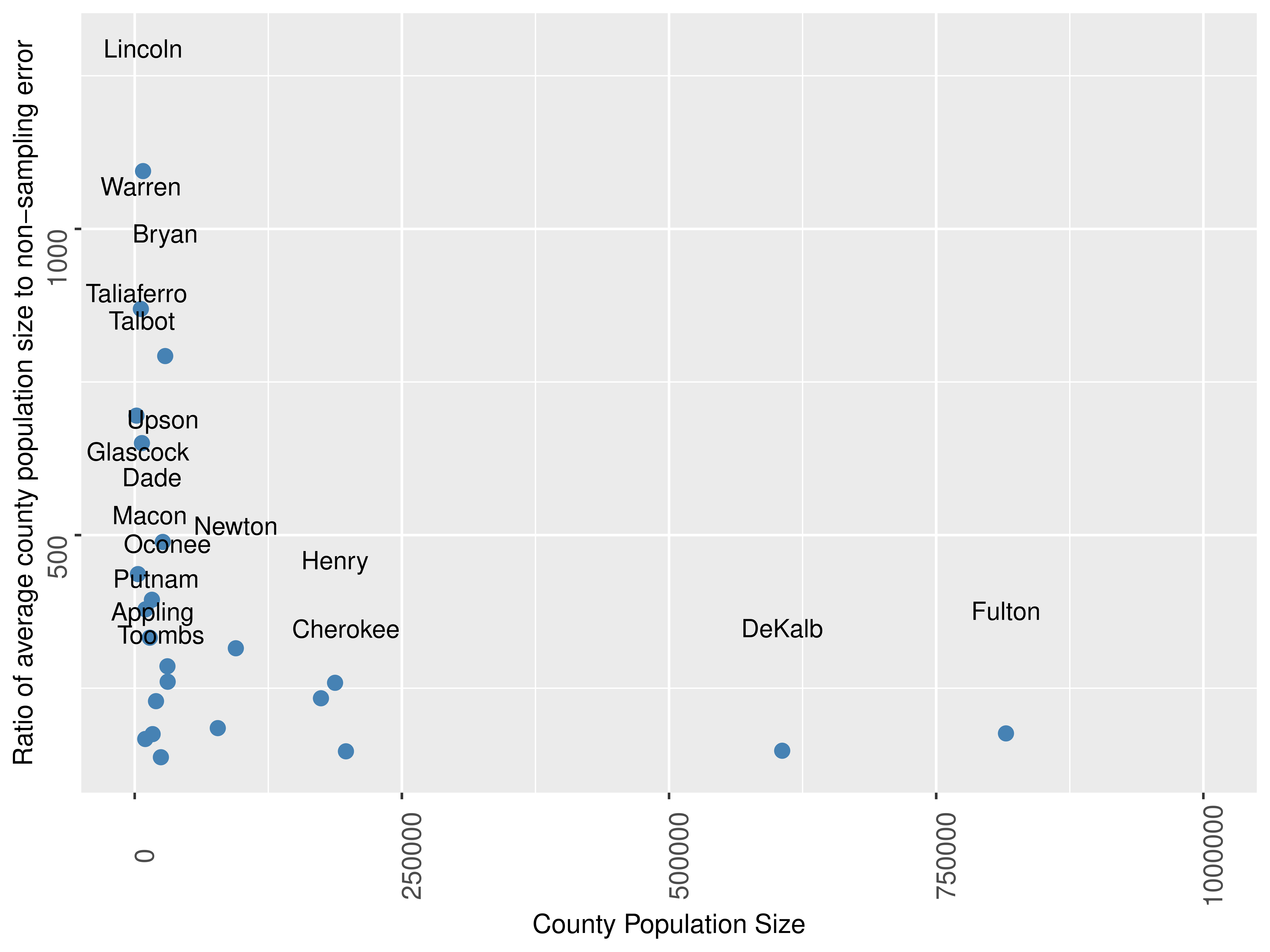}
 	\caption{Ratios of average county population size, across all years, to county-specific non-sampling error estimates for selected counties (n=23) in Georgia. County-specific ratios (y-axis) are plotted against average county population size (x-axis).}
 	\label{fig:nonsampratio}
 \end{figure}

Figure \ref{fig:nonsamperr} is a graphical representation of the posterior median, and $95\%$ CI estimates, of county specific levels of log-transformed non-sampling error, $log(\hat{\sigma}^2_{NS[c]})$ (shown in blue), for randomly selected counties in Georgia. These county specific estimates are plotted alongside log-transformed ACS reported sampling errors (red). The model-based estimates (blue), capture the uncertainty contributed by non-sampling errors, i.e., erroneous enumerations in survey counts. The county-specific decennial census population counts, and ACS reported population counts, inform the level of non-sampling errors via the data model assumptions shown in Section \ref{sec:datamod}. Comparison of county specific non-sampling errors shows the contrast between counties in regards to their levels of non-sampling errors. The smallest non-sampling error was attributed to Taliaferro County (a small population size county), with a median estimate of 1.69, $95\%CI = (-1.47, 2.89)$ (log scale). The largest non-sampling error was attributed to DeKalb County (a large population size county), with a median estimate of 5.56, $95\%CI = (5.50, 5.62)$ (log scale). Figure \ref{fig:nonsamperr} suggests that counties with higher population size, and more heterogeneity (Gwinnet, DeKalb, Cobb, and Fulton), suffer from substantially higher levels of non-sampling error. We performed a sensitivity analysis, shown in Figure \ref{fig:nonsampratio},  to assess if the ratio of county-average population counts to their estimated county-specific non-sampling error were higher for larger counties, in other words, are larger counties performing worse in estimating population size. We found no clear trend in ratios across counties indicating that larger counties, although suffering from higher error, do not have higher relative error than smaller counties.

\subsection{County population size estimates and uncertainties}\label{estresults}
\begin{figure}[H]
 \centering
 	\includegraphics[width=0.8\textwidth]{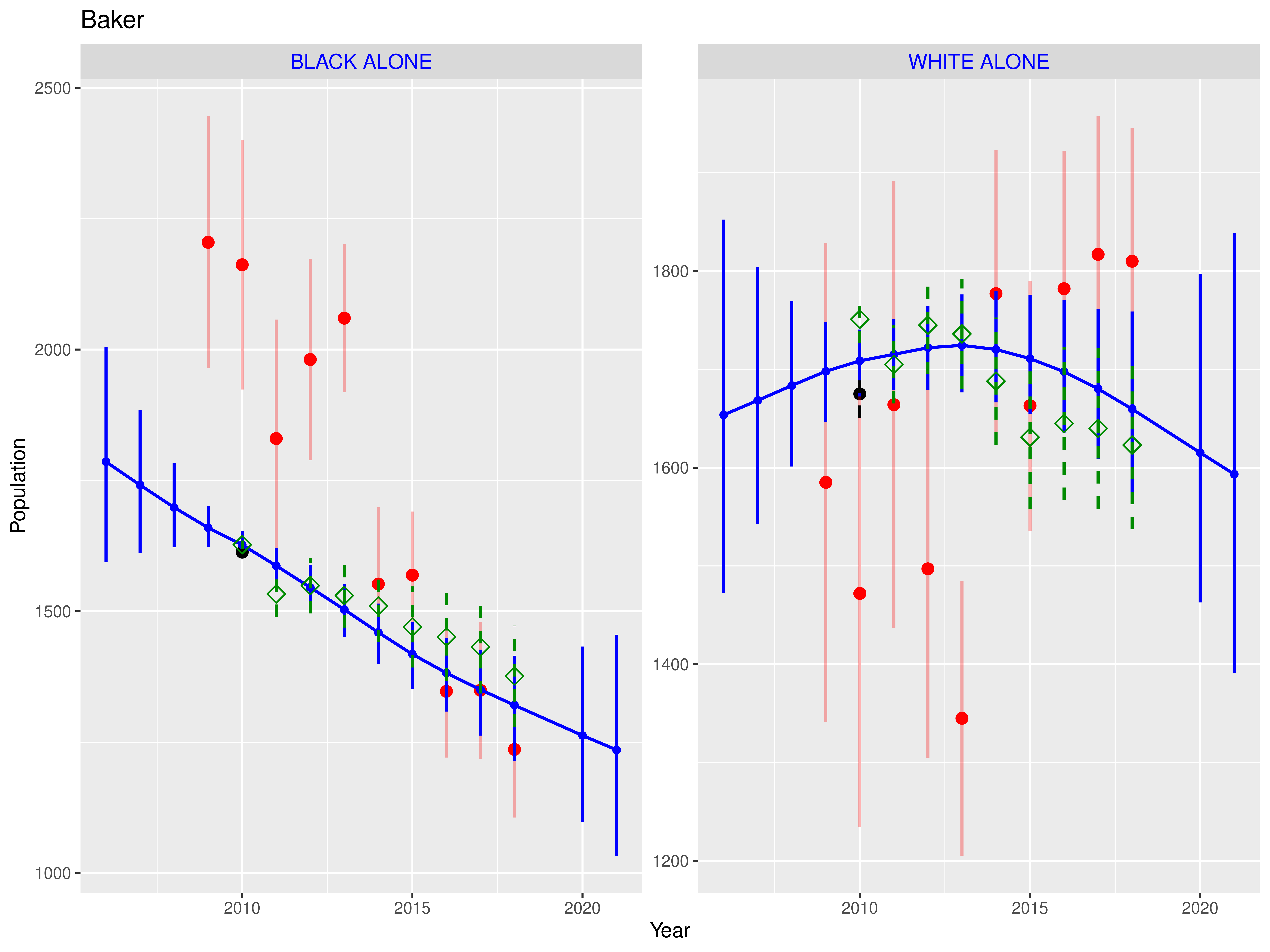}
 	\caption{Time trends of county-race specific posterior median estimates, and $95\%$ CIs, against observed data, for Baker County. B-Pop median estimates and $95\%$ credible intervals are shown in blue. Estimates are obtained for years 2005-2021. Red refers to ACS reported counts, black refers to decennial census counts, and green refers to PEP projection estimates. Data are shown for years 2009-2019 refer to the end years of ACS 5-year period estimates. }
 	\label{fig:baker}
 \end{figure}

\begin{sidewaysfigure}[ht]
    \includegraphics[scale=0.14]{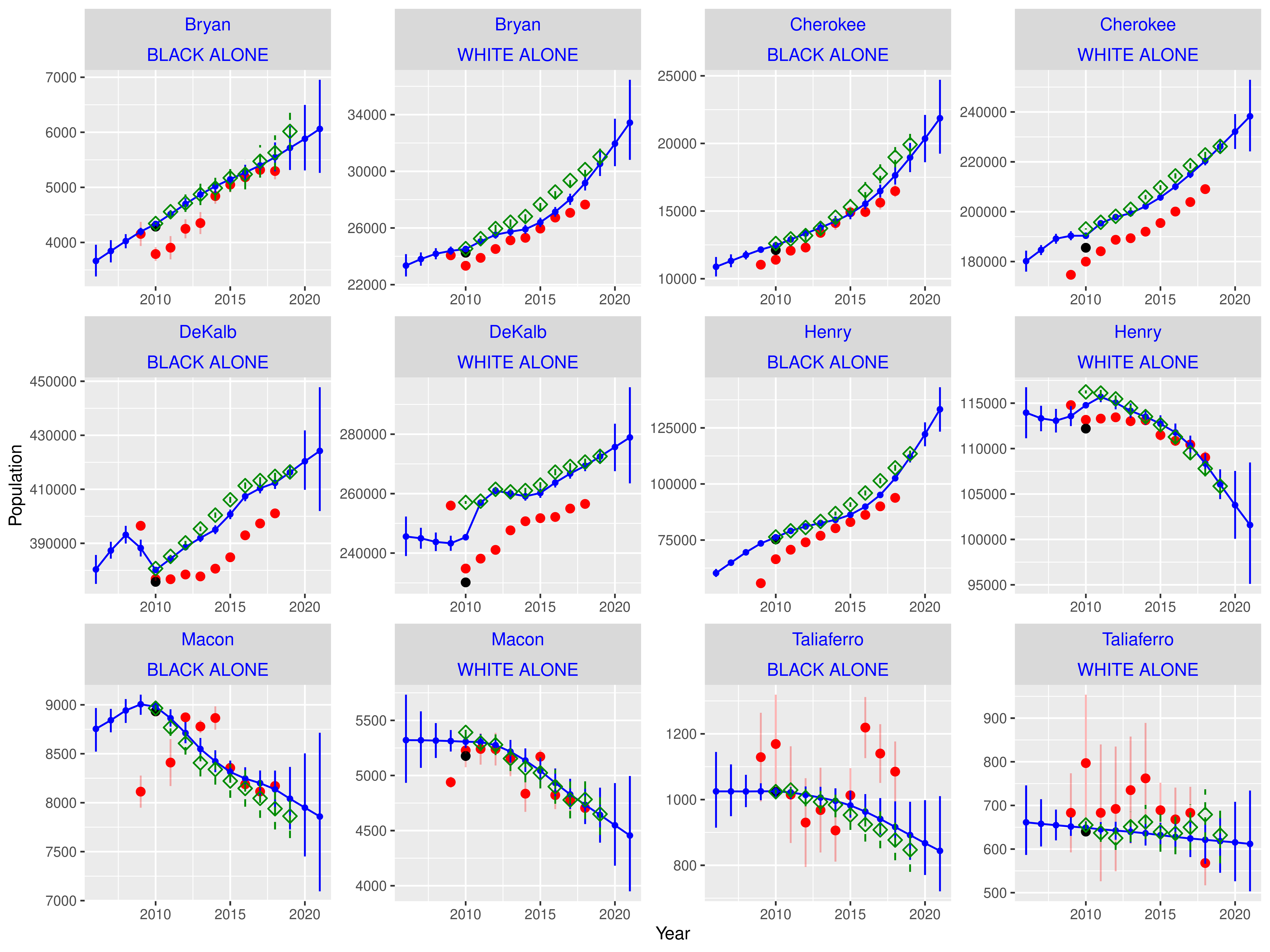}
   \caption{Illustration of B-Pop model data and county estimates for selected counties (Bryan, Cherokee, DeKalb, Henry, Macon, Taliaferro) stratified by race (White only, Black only). Parameters plotted are estimated population counts. The plots include: 1. observed ACS data with associated observation-based 95\% error intervals (red), 2. posterior estimates with 95\% credible intervals (blue), 3. Decennial census population counts (black), and 4. PEP population projections (green).}
    \label{fig:subplot}
\end{sidewaysfigure}

\begin{sidewaysfigure}[ht]
    \includegraphics[scale=0.14]{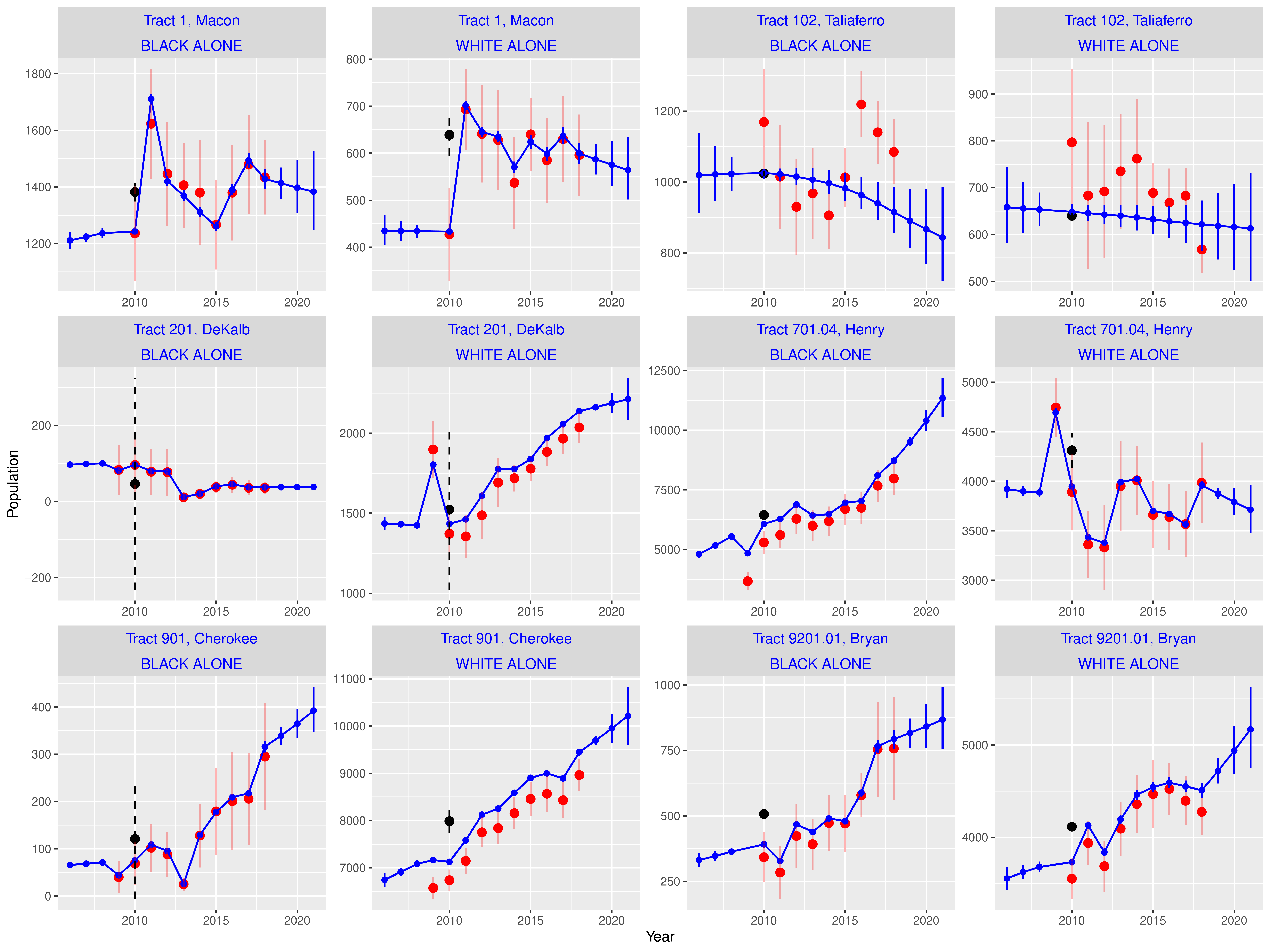}
   \caption{Illustration of B-Pop model data and tract estimates for selected counties (Bryan, Cherokee, DeKalb, Henry, Macon, Taliaferro) stratified by race (White Only, Black Only). Parameters plotted are estimated population counts. The plots include: 1. observed ACS data with associated observation-based 95\% error intervals (red), 2. posterior estimates with 95\% credible intervals (blue), and 3. Decennial census population counts (black).}
    \label{fig:longplot}
\end{sidewaysfigure}

Population estimates are shown for selected counties in Figures \ref{fig:baker} to \ref{fig:longplot}. Figure \ref{fig:baker} shows county-year-race stratified estimates (posterior medians and $95\%$ credible intervals) for Baker County. The figure shows in cases where the decennial census, PEP, and ACS are not consistent, there is heavy influence by the decennial census, and PEP reported populations. For example, in the Black only population estimate, we see shrinkage of the B-Pop estimate (blue) to values between the decennial census count (black), and PEP reported population (green). In contrast, outside decennial years, B-Pop estimates follow PEP observed trends more closely compared to ACS. In Section \ref{sec:datamod}, we describe our assumption for the variance parameters, in which we assume the decennial years suffer the least amount of error, i.e., the decennial years suffer from non-sampling error only. Therefore, with smaller uncertainty surrounding decennial counts, B-Pop estimates in the decennial year are heavily data-driven, shrinking more towards decennial data as compared to other sources (ACS and PEP reported counts).

Figure \ref{fig:subplot} shows county-year-race stratified estimates (posterior medians) of the population sizes from B-Pop for selected counties, with each source of population data overlaid. The figure shows that county-level race stratified population counts may differ substantially between ACS (red), PEP (green), and decennial data (black). For example, in very small counties, such as Taliaferro, the plots show much larger relative differences, ratio of population difference to ACS population estimate,  compared to the moderate and larger  counties. Taliaferro shows the largest relative difference of 0.21 for year 2014 in the black only population. Not accounting for population size, DeKalb county has the highest absolute difference of 19,920 in year 2012 for the White only population.

The B-Pop estimates (blue) closely follow PEP reported data trends, and the rate of change in B-Pop estimates follows closely to the observed rate of change in the PEP estimates, suggesting that B-Pop posterior 1-year population estimates are robust to deviating information given by ACS period estimates, and decennial cross-sectional estimates. This property is desirable because our aim is to estimate 1-year county-year-race specific true population estimates. Therefore, we expect our estimates to be closely related to data that capture 1-year population changes. Additionally, ACS population estimates consist of smoothed averages for the preceding five years, as such increases in population sizes will lag 1-year estimates. The plots show increased posterior uncertainty, surrounding population estimates, for smaller counties, such as that in Taliaferro, and Macon, where we expect higher sampling errors. We expect higher non-sampling error in larger counties, as shown in Figure \ref{fig:nonsamperr}. 
In addition, the predicted 2020 and 2021 population estimates show an increase in associated uncertainties, compared to earlier years, which is expected due to lack of direct data to inform these estimates.

Figure \ref{fig:longplot} shows tract-race specific 1-year estimates of the population for selected counties. Tract estimates are derived by applying ACS tract proportions to the county-year-race B-Pop estimates, shown in Section \ref{sec:dat}. Notably, the uncertainty, surrounding B-Pop tract level estimates, is derived from propagating uncertainty from the county-specific estimates. By using posterior samples of county population estimates, we propagate uncertainty into our tract level estimates. We acknowledge that this may not be a realistic assumption, uncertainty related to ACS proportions is not accounted for, and see propagation of uncertainty as a item for further research. For tracts within Bryan, Cherokee, DeKalb, and Macon counties, the true population estimates follow closely to ACS 5-year census tract-level estimates.

\clearpage

\subsection{Validation Results}
 The B-Pop model performs well in out-of-sample validation exercises, shown in  Table~\ref{tab:valid}.  We summarized the difference between model-based predictions, and both, PEP and ACS reported county population sizes.
 Median (MDE and MAE) and mean errors (ME and MSE) are close to 0, suggesting the model is well calibrated.  However, we see smaller errors associated with PEP data, as compared to ACS data,  median errors (MDE) are 0.0002 and -0.0013, and median absolute errors (MAE) are 0.0009 and 0.003, respectively. Mean squared error is similar between PEP and ACS associated errors, 0.0021 and 0.002, respectively. Lastly, the coverage of the 95\% prediction intervals is around 95\% for PEP reported population sizes with tail probabilities of (0.025, 0.021), but shows biased coverage when compared to ACS reported population sizes with larger tail probabilities of  (0.05, 0.08). This is due to inconsistencies between PEP 1-year reported counts and ACS 5-year reported counts. The larger differences between ACS and PEP reported data result in improved 95\% coverage for PEP reported population sizes.

\begin{table}[H]
\small
\centering
\resizebox{\textwidth}{!}{\begin{tabular}{|p{2.5cm} |p{2.2cm} |p{1.5cm}|p{1.4cm} p{1.2cm}|p{1.2cm} p{1.2cm}|p{1.2cm} p{1.2cm}|  }
\hline
\multicolumn{9}{|c|}{\textbf{{Error in county-year-race log population counts}}} \\
\hline
\textbf{Validation} & \textbf{Observed Data} & \textbf{N left-out} &  \multicolumn{2}{|c|}{\textbf{Median Errors} } & \multicolumn{2}{|c|}{\textbf{Mean Errors $(\%)$}} &  \multicolumn{2}{|c|}{\textbf{outside $95\%$ PI}}  \\
& &  &MDE & MAE & ME & MSE & Prop Below & Prop Above\\
  \hline
Leave-out years & PEP data & 1272  &0.0002 & 0.0009  &  0.0022 &  0.0021  & 0.025  & 0.021 \\
2016-2019& ACS data & 954 & -0.0013 & 0.003 & -0.009 & 0.002 & 0.05 & 0.08\\
 \hline
\end{tabular}}
\caption{Validation results. The outcome measures are: median error (MdE), median absolute error (MAE), mean error (ME), mean squared error (MSE), as well as the $\%$ of left-out observations below and above their respective $95\%$ prediction intervals (PI) based on the training set. 
}
\label{tab:valid}
\end{table}
We further illustrate out-of-sample validation results for selected counties in Appendix D.

\subsection{Comparison of population size estimates between USCB data sources and B-Pop}
\begin{figure}[H]
 \centering
 	\includegraphics[width=0.8\textwidth]{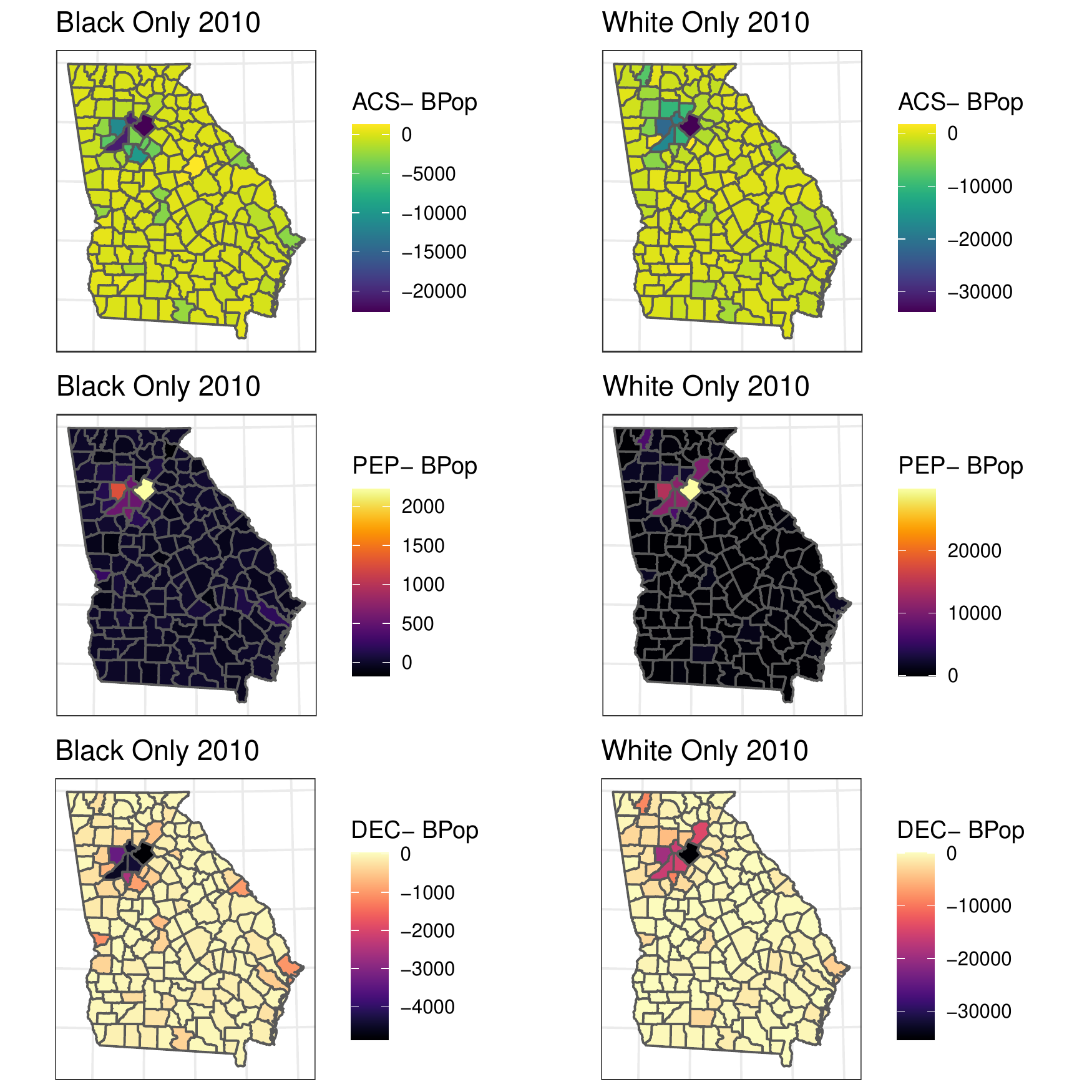}
 	\caption{Mapped differences in county population size estimates in 2010, broken down by race group. Differences were derived between ACS estimates and B-Pop, PEP and B-Pop, and Decennial census (DEC) and B-Pop. }
 	\label{fig:2010map}
 \end{figure}

\begin{figure}[H]
 \centering
 	\includegraphics[width=0.8\textwidth]{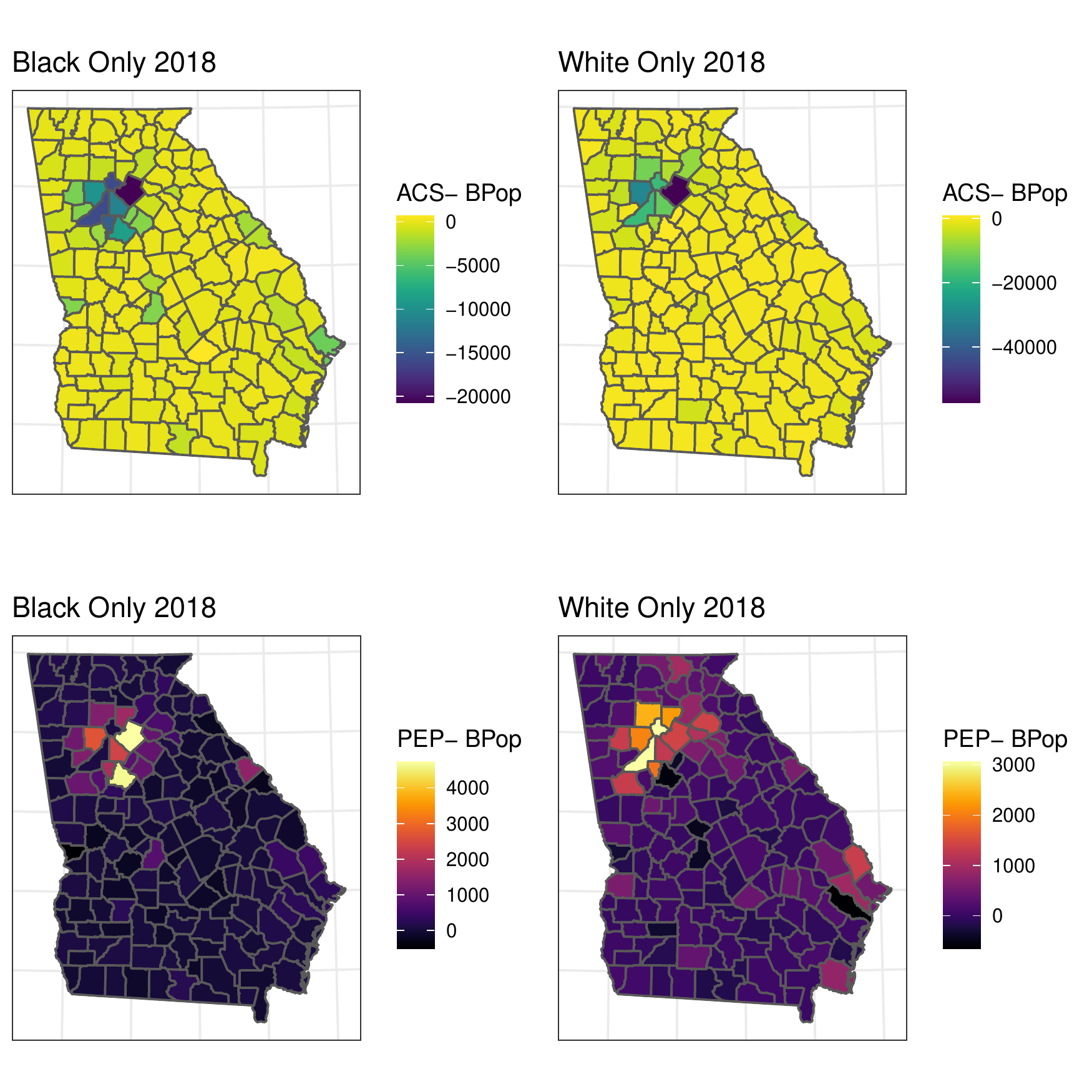}
 	\caption{Mapped differences in county population size estimates in 2018, broken down by race group. Differences were derived between ACS estimates and B-Pop, and PEP and B-Pop.}
 	\label{fig:2018map}
 \end{figure}

To compare B-Pop population size estimates with those of ACS, PEP, and Decennial census, Figures \ref{fig:2010map} and \ref{fig:2018map} mapped differences in county-specific estimates for both Black only and White only populations, for the years 2010 and 2018. Comparison of ACS to B-Pop in 2010 (Figure \ref{fig:2010map})  shows similar patterns across both Black only and White only groups, in which the highest absolute difference occurs in the most densely populated counties. The largest absolute difference was found to be for Gwinnett county's white population  with the absolute difference to be 33,740 and 57,302 individuals, for 2010 and 2018, respectively. 
Comparison of PEP to B-Pop in 2010 shows the more populated areas have higher differences in estimates. However, the differences between PEP and B-Pop are lessened compared to those of ACS or the Decennial census differences. The largest difference was also found to be in Gwinnett county's white population in 2010, equal to 29,860 individuals. The differences between B-Pop and PEP are greatly reduced in 2018, but show similar pattern to the difference between 2010 ACS and BPop estimates.

\begin{figure}[H]
 \centering
 	\includegraphics[width=0.8\textwidth]{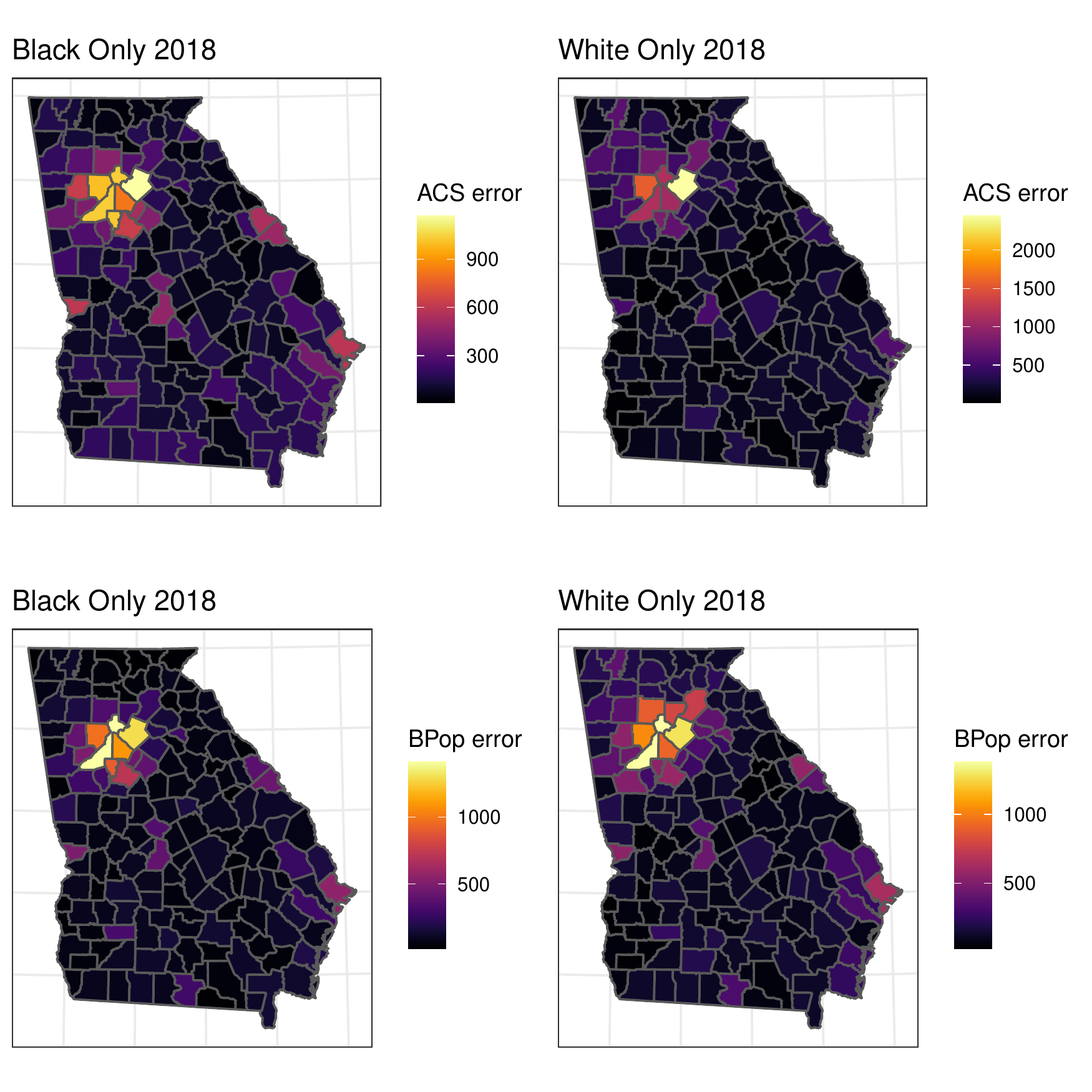}
 	\caption{Mapped county-specific estimates of errors for the year 2018, broken down by data source (ACS and B-Pop), and broken down by race.}
 	\label{fig:err2018}
 \end{figure}

Figure \ref{fig:err2018} maps county-specific 2018 error estimates broken down by data source (ACS versus B-Pop), and by race.  The results suggest that for both races, the more densely populated counties suffer from higher errors, with the errors being substantially higher using ACS data. Taking DeKalb county as an example, ACS sampling error is reported at $sd =1069.67$, while the total posterior error from B-Pop is $sd = 99.73$. The reduction in uncertainty can be attributed to the other data sources used to inform posterior estimation. The inclusion of PEP, and decennial census data, results in lower uncertainty surrounding B-Pop posterior estimates of the county-specific population sizes. Similar results are obtained when we consider race-specific population estimates.

\section{Discussion}
In this paper, we present Bayesian hierarchical model to fuse population count data from multiple USCB programs, each with different strengths and limitations, in a principled fashion to create a single set of small-area population estimates and uncertainties that can be used as denominators in the creation of rates for pubic health research. Characterizing data source specific errors is challenging because of the limited data availability from the decennial census, and lacking information on non-sampling errors associated with decennial census and PEP data. However, it is necessary to describe, and assess, data source specific errors in order to: (1) accurately understand the additional uncertainty introduced via ACS sampling design, and (2) obtain small area population estimates that incorporate data elements of variability.  We have described the B-Pop approach, which distinguishes between sampling, and non-sampling errors, and produces model-based estimates of the true county-year-race population counts, and better accounts for differing levels of uncertainty. 
B-Pop incorporates the differing data methodologies, within our model, to account for ACS sampling design, and decennial census non-sampling bias. We obtain county-year-race specific estimates of the true population size as well as tract-year-race specific estimates, responding to a need for up to date small area annual population estimates for smaller sub-populations.
To the best of our knowledge, no previous work has incorporated both data methodologies and data source specific errors to model true U.S. small area population counts in the context of multiple data sources and data sparsity. 

Model results suggest that the B-Pop model is well calibrated, and produces data-driven population estimates, with model-based estimates of non-sampling error. We illustrate how our proposed model produces estimates that capture population time trends within county, and account for the decrease in uncertainty surrounding decennial data. Higher uncertainty levels are expected for counties with smaller population sizes. However, in these smaller and highly variable counties, B-Pop produced estimates that proficiently captured the variable time trends. 

Limitations to our study relate to sparse or unavailable data and information needed to directly inform estimates of non-sampling errors. To overcome this challenge, strong modeling assumptions were made, and results may be sensitive to model form. Additionally, for simplification, our analysis included White only and Black only race strata, but does not include other race groups.

This work makes several contributions. First, we introduce a model that can account for distinct errors associated with different sources of USCB population data. In other words, we account for the increase in uncertainty due to the ACS sampling design compared to that of decennial census data. We estimate small area county-race-year specific population size, accounting for the different data generating processes of ACS, PEP, and decennial census data. We predict estimates to years without data. Lastly, we produce census tract-year-race estimates of true population counts, linking multiple data sources of small-area population data available from USCB. In doing so, we present a framework in which the population data, provided by USCB, can be combined together to learn about true population size.
Although we applied our model to 159 counties in Georgia, this approach can be applied to other counties within the U.S. to obtain estimates of county-year-sub population counts and non-sampling errors. This approach may also be extended to countries with multiple sources of population data. Future work will consist of further exploration of how to improve data-source specific uncertainties, specifically in the use of PEP data with additional considerations for vital statistics errors. Future work will also include comparison of USCB population data for years 2020 and 2021 as they are to become available shortly. More information is needed to understand and model uncertainty in particularly smaller counties to accurately capture time trends, and to allow for more precise predictions of county-year-sub population trends. Foremost, within the larger framework of small area estimation of disease and/or mortality rates, further research will entail answering the question of how to incorporate this information to accurately estimate disease and/or mortality rates while incorporating these estimates of uncertainty.

\clearpage
\section*{References}
Bradley, J. Wikle, C., and Holan, S. (2015). Saptio-temporal change of support with application to American Community Survey multi-year period estimates.arXiv.\\
Carroll, R.j., Ruppert, D., Stefanski, L.A., and Crainiceanu, C.M. (2006). Measurement Error in Nonlinear Models. Chapman \& Hall/CRC, second edition.\\
Chen, C., Wakefield, J., and Lumley, T. (2014). The use of sampling weights in Bayesian hierarchical models for small area estimation. Spatial and Spatio-temporal Epidemiology, 11:33-43.\\
Mercer, L. Wakefield, J., Checn, C., and Lumley, T. (2014). A comparison of spatial smoothing methods for small area estimation with sampling weights. Spatial Statistics, 8:69-85.\\
Mercer, L., Wakefield, J., Pantazis, A., Lutambi, A.M., Msanja, H., and Clark, S. (2015). Space-Time Smoothing of Complex Survey Data: Small Area Estimation for Child Mortality. Annals of Applied Statistics, 9(4).188901905.\\
Nethery, R., Rushovich, T., Peterson, E., Chen, J., Waterman, P., Krieger, N., Waller, L., and Coull, B. (2021). Comparing the performance of three census tract denominator sources for real-time disease incidence modeling: US decennial census, American Community Survey, and WorldPop. Health and Place.\\
Plummer, M. (2017). JAGS: A program for analysis of Bayesian graphical models using Gibbs sampling. In proceedings of the 3rd International Workshop on Distributed Statistical Computing.\\
Population Estimation Program. (2019). Methodology for the United States Population Estimates: Vintage 2019. \\
Preston, S., Heuveline, P., and Guillot, M. (2000). Demography: Measuring and Modeling Population Processes. Wiley Blackwell, first edition.\\
Shumway, R.H. and Stoffer, D.S. (2006). Time Series Analysis and its Applications. New York Springer, sixth edition.\\
Spielman, S. and Folch, D. (2015). Reducing Uncertainty in the American Community Survey through Data-Driven Regionalization. PLos ONE, 10(2):1-21.\\
Spielman, S., Folch, D., and Nagle, N. (2014). Patterns and causes of uncertainty in the American Community Survey multiyear estimates. In Proceedings of the section of survey research methods. Alexandria, VA: American Statistical Associations, pages 3011-3017.\\
United State Census Bureau (2012). Decennial census: Complete technical documentation.\\
U.S. Census Bureau (2009). A compass for understanding and using American Community Survey data: What researchers need to know.\\
U.S. Census Bureau (2014). The American Community Survey. Design and Methodology. U.S. Government Printing Office.\\
U.S. Census Bureau (2014). American Community Survey: Design and Methodology Chapter 11: Weighting and Estimation.\\
U.S. Census Bureau (2018). Understanding and Using American Community Survey Data: What all data users need to know.\\
U.S. Census Bureau: CCM (2012). Estimates of undercount and overcount in 2010 census. \\
Walker, K. (2020). Tidycensus: Load US census boundary and attribute data as `tidyverse'. R package version 0.9.9.2.

 \clearpage
\section{Appendix A: Notation Table}\label{sec:app}
\resizebox{\textwidth}{!}{
\begin{tabular}{ |c|c|} 
 \hline
 \textbf{Parameter Notation} & \textbf{Description} \\ 
 \hline
 \hline
\multicolumn{2}{c}{\textcolor{blue}{Data Quantities}}\\
 \hline
 \hline
$n_{c,t,r}^{(ACS)}$ & ACS reported population size for county $c$, year $t$, race $r$ \\
\hline
$s_{c,t,r}^2$ & ACS reported data variance (sd) for county $c$, year $t$, race $r$ \\
\hline
$n_{c,r}^{(census)}$ & Decennial census reported population size for county $c$,race $r$ \\
\hline
$n_{c,t,r}^{(PEP)}$ & PEP reported population size for county $c$, year $t$, race $r$ \\
\hline
$p_{c,t,r}^{(PEP)}$ & PEP reported proportion of annual population size out of aggregate population across 5-year intervals.\\
\hline
$p_{c[g],r,t}^{(ACS)}$ & Ratio of ACS reported population for tract $g$ in county $c$, year $t$, race $r$ to the sum of the population across all tracts within county $c$\\
 \hline
 \hline
\multicolumn{2}{c}{\textcolor{blue}{Estimates Quantities}}\\
 \hline
 \hline
$\gamma_{c,t,r}$ & True population size for county $c$, year $t$, race $r$ \\
\hline
$\tilde{\gamma}_{c,r}$ & Weighted average of true population size for county $c$,race $r$, for years $t_1$ to $t_1 +5$ \\
\hline
$\gamma^{(s)}_{c[g],r,t}$ & Sample $(s)$ of true population size for tract $g$ within county $c$, year $t$, race $r$ \\
\hline
$\eta_{c,t,r}$ & True log-population size for county $c$, year $t$, race $r$ \\
\hline
$\eta_{c,r}$ & True population level across county $c$,  race $r$ \\
\hline
$\eta_{c}$ & True population level across county $c$ \\
\hline
$\eta_{global}$ & Global level of true population size \\
\hline
$\delta_{c,t,r}$ & Estimated annual rate of change between log population counts for county $c$, race $r$, and years $t$ and $t-1$.\\
\hline
$\sigma^2_{c,t,r}$ & total variance for county $c$, year $t$, race $r$ \\
\hline
$\sigma^2_{NS[c,r]}$ & total non-sampling variance for county $c$, race $r$ \\
\hline
$\phi$ & County-race variability in true population size \\
\hline
$\omega$ & Across county variability in true population size \\
\hline
$\sigma^2_{\eta}$ & Across county-race-year variability in true population size outside the decennial year \\
\hline
$\sigma^2_{ref}$ & Across county-race variability in true population size within the decennial year.\\
\hline
$\sigma^2_{\delta}$ & Across county-race-year variability in the population annual rates of change.\\
\hline
$\xi_{c}$ & Decennial data net under-count adjustment for county $c$ \\
 \hline
\end{tabular}}

\section{Appendix B: Model Jags Code}
\begin{lstlisting}[language=R]
  model {


##################################################################
# Process model 
##################################################################
for(c in 1:C){
for(r in 1:R){

### RW (2) for log-transformed population parameters

#Reference year
eta.ctr[c,tref,r] ~ dnorm(eta.cr[c,r],  tau.ref)
eta.ctr[c,tref-1,r] ~ dnorm(eta.cr[c,r],  tau.ref)

#Forward projection from reference year
for(tindex in (tref+1):nyears){
eta.ctr[c,tindex,r] ~ dnorm(2 *eta.ctr[c,tindex-1,r] - eta.ctr[c,tindex-2,r],  tau)
}

#Backward projection from reference year
for(tindex in 3:tref){
eta.ctr[c,tindex-2,r] ~ dnorm(2 *eta.ctr[c,tindex-1,r] - eta.ctr[c,tindex,r],  tau)
}


## Errors for annual population change  d_[c,t,r]
 for(t in 1:nyears){
  tau_r.ctr[c,t,r] = 1/(sigma_r.ctr[c,t,r]^2)
  sigma_r.ctr[c,t,r] ~ dnorm(sigma_r.cr[c,r], 0.01)T(1,1000)
  }

## PEP errors:
    #In decennial year = nonsampling error
    #Outside decennial year = nonsampling error + cumulative sum of var(d.ctr's)

    tau_pep.ctr[c,tref,r] = 1/sigma_nonsamp.cr[c,r]^2

    for(t in (tref+1):(nyears-2)){
    tau_pep.ctr[c,t,r] = 1/sigmasq_pep.ctr[c,t,r]
    sigmasq_pep.ctr[c,t,r] = sigma_nonsamp.cr[c,r]^2 + sum(sigma_r.ctr[c,(tref+1):t,r]^2)
    }



##################################################################
# Hierarchical parameters (c-r) and (c) parameters
##################################################################

  #log-population level [c,r]
  eta.cr[c,r] ~ dnorm(eta.c[c], phiinv)

  #non-sampling errors [c,r]
  tau_nonsamp.cr[c,r] = 1/(sigma_nonsamp.cr[c,r] ^2)
  sigma_nonsamp.cr[c,r] ~ dnorm(sigma_nonsamp.c[c],0.01)T(0,)
 

  #annual pop change error [c,r]
  sigma_r.cr[c,r]~ dnorm(sigma_r.c[c],0.01)T(0,)


  }#end R loop


  #log-population level [c]
  eta.c[c] ~ dnorm(etaglobal, omegainv)

  #annual pop change level [c]
  sigma_r.c[c] ~ dnorm(0,0.01)T(0,)


 #non-sampling error level [c]
  sigma_nonsamp.c[c] ~ dnorm(0,0.01)T(0,)

  # Under-reporting adjustment [c]
  xi.c[c] ~ dnorm(0.91, 1/(1.04^2))T(0,1)


  }#end C loop



##################################################################
# Hyper-parameters
##################################################################
    # global level of population
    etaglobal ~ dnorm(0, 0.001)

    #global correlation across years
    rho ~ dunif(0,1)

    # variance before/after the RW reference year
    tau <- 1/(sigma_eta^2)
    sigma_eta ~ dnorm(0, 0.1)T(0,)
   


    #variance in the RW reference year
    tau.ref<- 1/(rsigma_eta^2)
    rsigma_eta ~ dnorm(0, 0.1)T(0,)
 

    #across county and race variance in log(population) 
    phiinv <- 1/phi^2
    phi ~ dnorm(0, 0.1)T(0,)
  

   #across county variation in log(population)
    omegainv <- 1/omega^2
    omega ~ dnorm(0, 0.1)T(0,)
   
##################################################################
# Data Model
##################################################################

      for(c in 1:C){
      for(r in 1:R){


      #### PEP data model tau_pep = precision of nonsampling error + annual change associated errors

      for(t in tref:(nyears-2)){
      pep_pop.ctr[c,t,r] ~ dnorm(gamma.ctr[c,t,r], tau_pep.ctr[c,t,r])
      }


      ### Population change data model, tau_r = precision of annual change.
    for(t in (tref+1):(nyears-2)){
    d.ctr[c,t,r]~ dnorm(0,  tau_r.ctr[c,t,r])
    }

      ### Decennial census data model, tau_nonsamp  = nonsamp precision
      decpop_tref.cr[c,r] ~ dnorm(gamma.ctr[c,tref,r] * (1- (xi.c[c]/100)), 1/sigma_nonsamp.cr[c,r]^2)



      ###ACS datamodel
      
     acspop.cpr[c,1:p_periods,r] ~ dmnorm( tildegamma.cpr[c,1:p_periods,r], Tau.cppr[c,1:p_periods,1:p_periods,r])


      # ACS var-cov (error) matrix
      Tau.cppr[c,1:p_periods,1:p_periods,r] <- inverse(Sigma.cppr[c,1:p_periods, 1:p_periods,r])

      for(p in 1:p_periods){
       tildegamma.cpr[c,p,r] = sum(gamma.ctr[c,p:(p+4),r] %*% pep_props.ctr[c,p:(p+4),r])
      Sigma.cppr[c,p,p,r] = sigma_nonsamp.cr[c,r]^2 + s.cpr[c,p,r]^2

       for(s in (p+1):p_periods){
       Sigma.cppr[c,p,s,r] = rho * sqrt(sigma_nonsamp.cr[c,r]^2 + s.cpr[c,p,r]^2) * sqrt(sigma_nonsamp.cr[c,r]^2 + s.cpr[c,s,r]^2)
       Sigma.cppr[c,s,p,r] = Sigma.cppr[c,p,s,r]
      }
      }



for(t in 1:nyears){
  gamma.ctr[c,t,r] = exp(eta.ctr[c,t,r])
    }

    }#end R
} # end C


      #   ----------------------------------
  }  #end model
\end{lstlisting}

\section{Appendix C: Hyper-Parameter Plots}\label{sec:plotprior}
\begin{figure}[H]
 \centering
 	\includegraphics[width=0.9\textwidth]{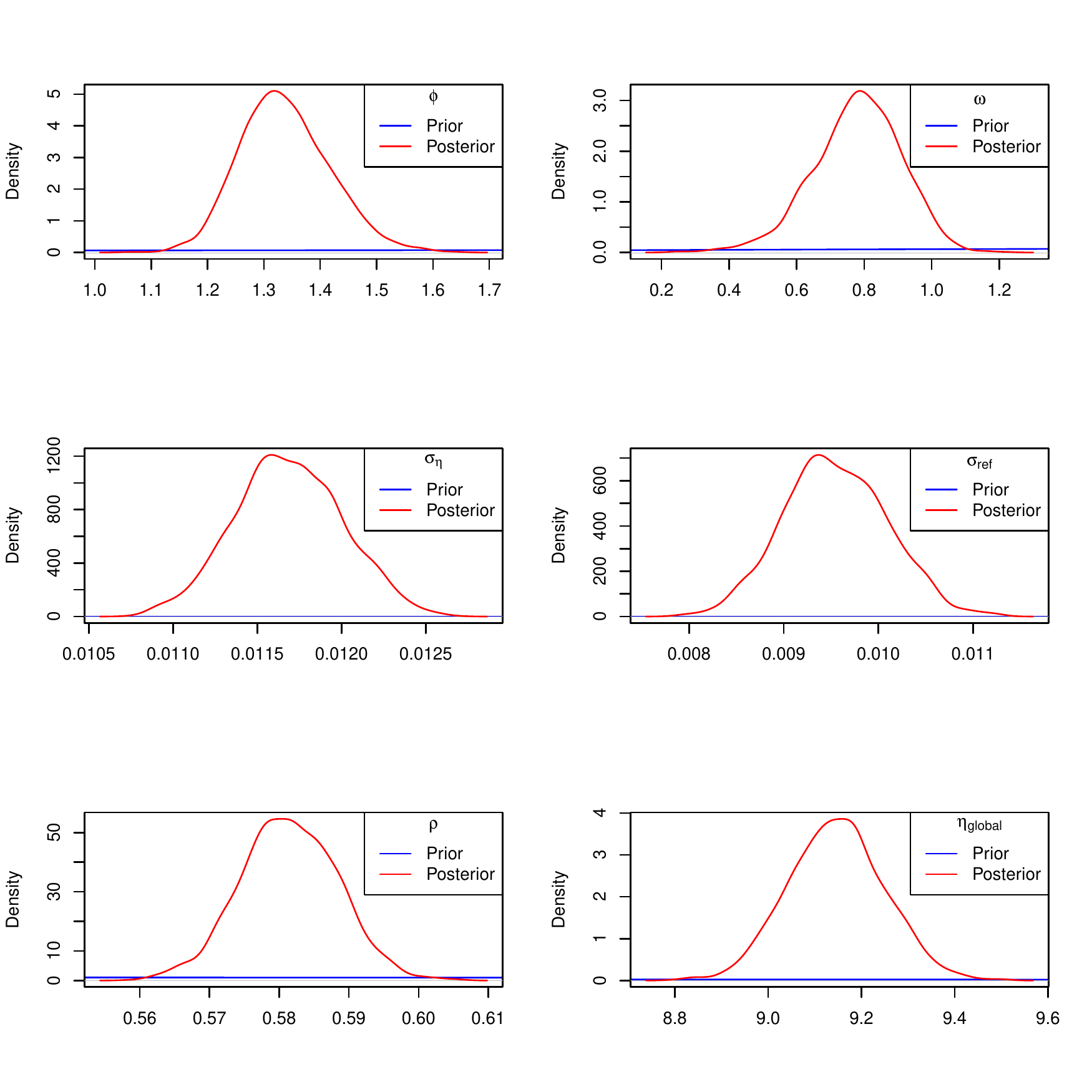}
 	\captionsetup[figure]{font=small,skip=0pt}
 	\caption{Illustration of posterior updating from prior distributions (blue) to the posterior distributions (red), for all model hyper-parameters. }
 	\label{fig:priors}
 \end{figure}

\noindent
We show plots of posterior distributions against prior distributions for all hyper-parameters, shown in Figure \ref{fig:priors}. Posterior distributions of the variance hyper-parameters $(\phi, \omega, \sigma_{\eta}, \sigma_{ref})$ are plotted against truncated normal distributions. The posterior distribution of the global population level, $\eta_{global}$, is plotted against a $N(0,100)$ prior distribution.  Figure \ref{fig:priors} suggests that there is substantial updating from prior to posterior distributions, which indicates that variance parameters are not heavily driven by their prior assumptions.

\section{Appendix D: Validation results for selected counties}\label{sec:vali}

\begin{figure}[H]
 \centering
 	\includegraphics[width=0.9\textwidth]{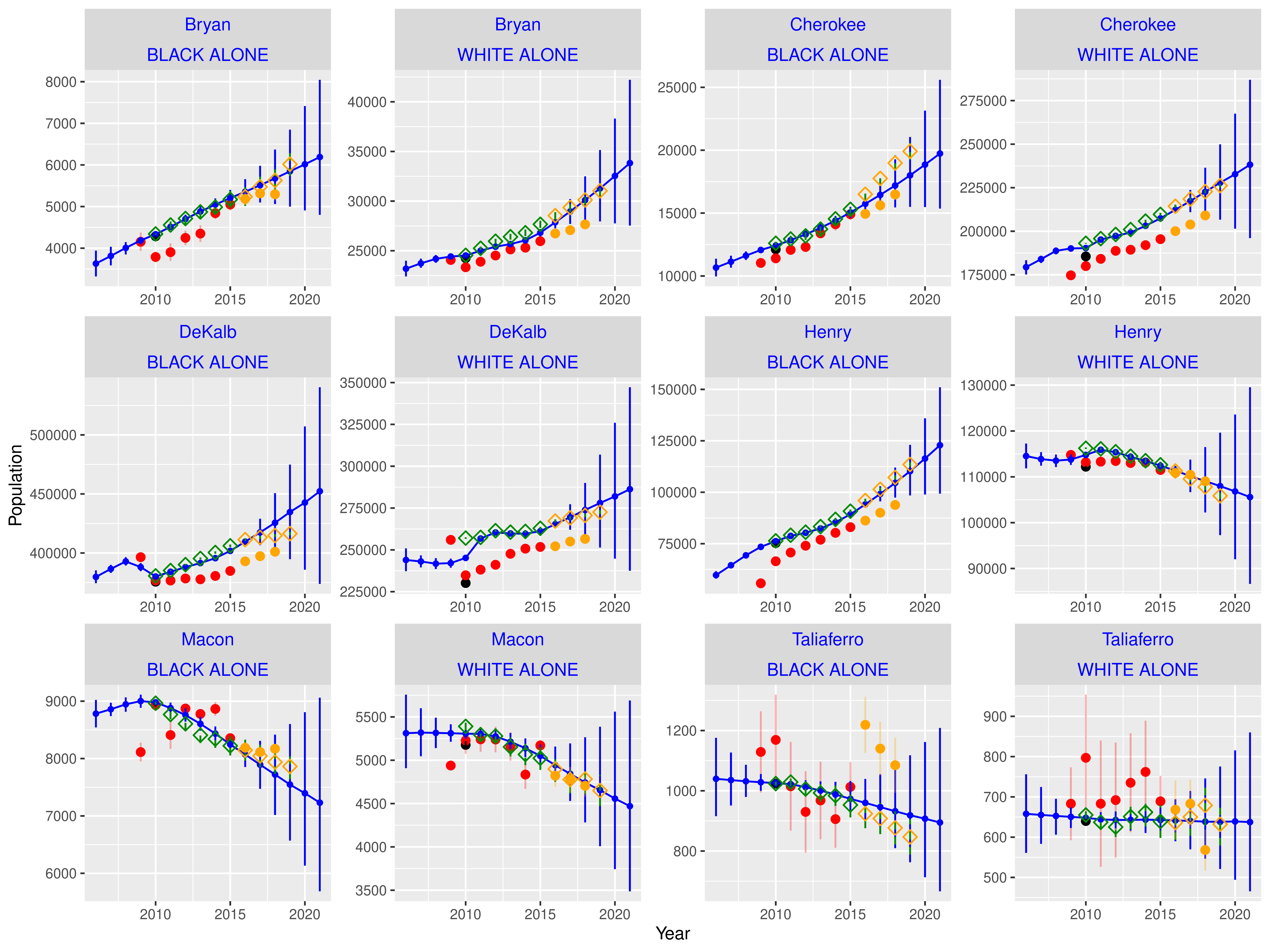}
 	\captionsetup[figure]{font=small,skip=0pt}
 	\caption{Validation results of B-Pop county estimates for selected counties (Bryan, Cherokee, Dekalb, Henry, Macon, Taliaferro) stratified by race (Black only, White only). Parameter estimates and associated $95\%$ CIs plotted in blue. Included ACS estimates and associated errors plotted in red. Included PEP estimates and associated errors plotted in green. Observed population counts, excluded for years 2014-2019, shown in orange (points for ACS excluded data, and diamonds for PEP excluded data).}
 	\label{fig:val}
 \end{figure}

Figure \ref{fig:val} illustrates validation results for selected counties. The left-out (excluded) observations are illustrated in orange for both PEP data (diamonds) and ACS data (points). We show that B-Pop predictions, for those years with left-out data, closely follow PEP projections in the case where PEP and ACS differ. This illustrates that B-Pop predicted estimates, for years 2014-2021, capture the observed trends well even when those data have been excluded. 

\end{document}